\documentstyle[preprint,aps,eqsecnum]{revtex}
\begin{document}
\draft
\title{Overscreening of magnetic impurities in
$d_{x^2-y^2}$ wave superconductors}
\author{Carlos R.~Cassanello$^{a,b}$ and Eduardo Fradkin$^{a}$}
\address
{Loomis Laboratory of Physics and Materials Research Laboratory$^{a,b}$\\
University of Illinois at Urbana-Champaign,
1110 W.Green St., Urbana, IL, 61801-3080\\
and Institut f{\"u}r Theoretische Physik$^{b}$ \\
Universit{\"a}t zu K{\"o}ln, Z{\"u}lpicher Str. 77,
D-50937, K{\"o}ln, Germany}
\bigskip
\maketitle

\begin{abstract}
We consider the screening of a magnetic impurity in a
$d_{x^2-y^2}$ wave superconductor.
The properties of the $d_{x^2-y^2}$ state lead to an unusual
behavior in the impurity magnetic susceptibility, the impurity specific
heat and in the quasiparticle phase shift which can be used to diagnose
the nature of the condensed state. We construct an
effective theory for this problem
and show that it is equivalent to a multichannel
(one per node) non-marginal Kondo problem with linear density of states
and coupling constant $J$.
There is a quantum phase transition from an unscreened impurity state to
an overscreened Kondo state
at a critical value $J_c$ which varies with $\Delta_0$, the
superconducting gap away from the nodes.
In the overscreened phase,
the impurity Fermi level $\epsilon_f$ and the amplitude $\Delta$ of
the ground state singlet  vanish at $J_c$ like
$\Delta_0 \exp(-{\rm const.~}/ \Delta)$ and $J-J_c$ respectively.
We derive the scaling laws for the susceptibility and specific heat in
the overscreened phase at low fields and temperatures.
\end{abstract}
\bigskip


\narrowtext

\section{Introduction}
\label{sec:intro}

The problem of magnetic impurities in d-wave superconductors has been
a subject of intense research~\cite{pines,wenger,balatsky,franz}. It is
known that in anisotropic superconductors, such as a $d_{x^2-y^2}$ state,
magnetic impurities act as pair breaking centers\cite{pair} and
hence reduce the amplitude of the condensate.
Experimentally, the main effect of these impurities is to reduce the
critical temperature $T_c$ of the superconducting state\cite{nmr-pines,nmr}.

Quantum mechanical fluctuations of magnetic impurities
give rise other important  effects particularly
when coupled to the fermionic quasiparticles of the
superconducting state. In a normal
metal these correlations lead to the Kondo screening of the impurity and
to the generation of dynamical energy scales such as the Kondo
temperature. In high temperature superconductors, the effects of the
magnetic impurities appear to depend significantly not only on the
nature of the impurity but also on where the magnetic impurity is
located. The conventional interpretation of the role of magnetic
impurities in high temperature superconductors relies, for the most
part, on the chemical differences of the impurities
(mainly $Zn$ and $Ni$) and on the actual location of the impurities on the
lattice relative to the $CuO$ planes. The main focus of recent work on this 
subject  
has focused on how much different impurities are able to depress
$T_c$\cite{nmr} and on the power laws that static (or clasical) impurities 
induce on 
low tempreture properties. 

In this paper we investigate the physics that results from the exchange coupling 
between isolated magnetic impurities and the quasiparticles of the 
superconducting
state. In particular we will be interested in finding out under what 
circumstances 
there is a Kondo-like 
dynamical screening of the magnetic impurity by the quasiparticles.
The  mechanism that we have in mind is analogous to the exchange
coupling between magnetic impurities and the electrons of a Fermi liquid
that causes the Kondo effect. However, unlike the Kondo
effect in metals, because the density of states of normal quasiparticles in
a d-wave superconductor vanishes at the Fermi energy, screening is absent
in perturbation theory and a critical exchange coupling between
the quasiparticles and the magnetic impurity is necessary for Kondo
screening
to take place. Thus, magnetic impurities which couple strongly to the
quasiparticles, such as $Zn$ which substitute for $Cu$ is the planes,
may actually be Kondo screened at very low temperatures while
$Ni$, which appears to couple more weakly, may not get to be in a Kondo
screening regime. Likewise, magnetic impurities on sites away from the
$CuO$ planes are weakly coupled and therefore are less likely to
undergo Kondo
screening. We will see below that the critical coupling (which we will
only estimate very roughly in this work) is controlled by $\Delta_0$,
the size of the gap away from the nodes at zero temperature and
tipically it is a fraction of $\Delta_0$.

The onset of Kondo screening at a critical  value of the exchange coupling
constant is a quantum critical point. We will show in this work that
the behavior of the magnetic impurity, both near  and beyond the phase
transition, has unique signatures which follow from the nature of the
condensate and  hence can be used to investigate its nature. Among its
most salient features are the temperature and magnetic field dependence
of the impurity magnetic
susceptibility and specific heat which exhibit strong deviations from Fermi
liquid behavior. We will also show  that the quasiparticle phase shift
exhibits a strong frequency dependence and a broad resonance and that
the structure of the quasiparticle  scattering matrix has
detailed information 
on the phases and signs associated with d-wave superconductivity. Thus,
the physics of magnetic impurities in a d-wave superconductor can be used to
diagnose the nature of the superconducting order. The purpose of this paper
is to describe these effects in detail.

The Kondo problem in metals~\cite{kondo} has been intensively and extensively
studied and it is by now very well understood. It is described in terms of 
 a {\it smooth} crossover from a marginally unstable fixed point at
zero exchange coupling to a stable Fermi liquid fixed point with a screend 
impurity
~\cite{pwa,nozieres,wilson}. The validity of this picture has been confirmed
by the exact solution by the Bethe ansatz\cite{andrei,wiegmann} and by
large-$N$ methods\cite{read}.
We will refer to this case as  to the {\it marginal} Kondo problem since
it bears a strong resemblance to a critical system at its low critical
dimension.

In a conventional s-wave superconductor the Kondo effect is suppressed by the 
formation 
of the superconducting 
gap, as shown by the classic theory of Abrikosov and Gorkov~\cite{abgkv}. 
However, in 
the case of a 
d-wave superconductor, there are quasiprticle states inside the superconducting 
gap 
which concentrate near the 
nodes of the order parameter. Although the density of states vanishes at the 
Fermi 
energy, for strong enough 
exchange coupling it may still be possible that isolated magnetic impurities may 
still 
be screend by the 
quasiparticles. The central idea of this work is that the way this screening 
happens 
may be used as a tool to 
study the superconducting order.

Impurities cause many different effects in superconductors. In the case of a 
d-wave 
symmetry, any sort of 
scattering breaks pairs\cite{pair} and, for instance,
static magnetic impurities produce quasiparticle bound states in the gap of the 
superconductor.
However, the binding energy of these states vanishes in the vicinity of
the nodes of the d-wave superconductor~\cite{balatsky,franz,shiba}.
Random potential scattering also leads to interesting 
effects, in particular close to the nodes of the superconductor where the 
density
of states (DOS) of quasiparticle states, which behave like Dirac fermions near 
the 
nodes, {\it vanishes} 
linearly with the
energy (measured from the Fermi energy).
It has also been shown that random potential scattering should generally
lead to a {\it finite} DOS at the Fermi energy (zero) for Dirac fermions in
random  potentials\cite{fisher,dirac,ludwig} and in d-wave superconductors
\cite{palee,wenger,hirshfeld1}. 
The precise behavior of the DOS appears to depend on how
many nodes are coupled and on what  channels  are
mixed by the scattering processes\cite{wenger,ludwig,hirshfeld2,hirshfeld3}.
However, if the superconductor is sufficiently clean, the effective
DOS induced by the disorder is {\it exponentially} small\cite{fisher,dirac}
and its effects can be neglected.
Notice, however, that the combined effects of Kondo
screening and random scattering is a problem that is still not undertood, even 
in 
metals.
~\cite{kotliar,phillips}.

The problem of a quantum
magnetic impurity coupled to fermions with a vanishing DOS at the Fermi
energy was first discussed by Withoff and Fradkin\cite{withoff}. 
In contrast with the Fermi liquid which has a
finite DOS at the Fermi energy , Kondo screening of the magnetic impurity
could only happen for values of the exchange coupling constant $J$ larger
than some critical value. If the DOS vanishes with
a power of the energy with exponent $r$, the fixed point at $J=0$ is
stable for $r>0$ and a new unstable fixed point appears at $J_c>0$
signaling a quantum phase transition. We will refer to this problem as the
non-marginal Kondo problem and $r$ measures the deviation from marginality.
Using a close analogy with the theory of critical phenomena, Withoff
and Fradkin developed a large-$N$ theory
for this problem and found that the essential singularity of the Kondo
temperature is replaced by power law singularities determined by the
DOS exponent $r$. However, in that work only the regime $0<r\leq
{\frac{1}{2}}$ was considered. The behavior of these systems for small
$r$ was also exlored by Chen and Jayaprakash\cite{jay} and by
Ingersent\cite{ingersent} who used a generalization of Wilson's
numerical renormalization group (RG) for this problem. 
The case of interest for a d-wave superconductor is $r=1$ which turns out to be 
special 
in several ways.
In ref.~\cite{paperII} we discussed recently the closely related problem
of a magnetic impurity in a flux phase which is also an example of an $r=1$
system. In this paper we show that these two problems can
be mapped into each other in spite of the fact that spin fluctuations do
break Cooper pairs.

One aspect of the physics of magnetic impurities in a d-wave superconductor that 
we 
will not consider here are the effects of the depression (and/or actual
vanishing of) of the d-wave  order parameter near the impurity site. This effect 
is 
independent of the spin and it actually static. In any event, the vanishing of 
the 
condensate at the impurity site leads to terms that do not conserve fermion 
number in 
the effective Hamiltonian and hence may lead to important (spin-independent) 
effects. 
This is a conceptually important problem and it will be addressed 
elsewhere\cite{paper4}. 
A self-consistent calculation based on the BCS approximation can be 
found in ref.~\cite{hirshfeld3}.

In this paper we will make use of a very simple model of the quasiparticle
dynamics in a $d_{x^2-y^2}$ superconductor\cite{scalapino,schrieffer}.
We will use the fact, strongly supported by the corner-junction interference
experiments\cite{wollman} as well as by
angle resolved photoemission spectroscopy (ARPES)\cite{shen,campuzano}
that the high temperature superconductors have a $d_{x^2-y^2}$ condensate
with four symmetrically arranged nodes where the quasiparticle gap
vanishes.
The first evidence that the gap vanishes at the $d_{x^2-y2}$ nodal line
was reported by Shen {\it et.al.,\/}\cite{shen}.
Ding {\it et.al.,\/}\cite{campuzano} reported measurements of the
momentum dependence of the superconductor gap in
${\rm Bi}_2{\rm Sr}_2{\rm CaCu}_2
{\rm O}_{8+x}$ consistent with a gap function of the form
$\cos(k_x) - \cos(k_y)$, as expected for a $d$-wave order parameter.
Interestingly enough, in underdoped systems, photoemission
supports the  idea that the gap may survive  through a large
range of temperature into the normal state~\cite{shen,campuzano}.
It is also well established that the high temperature superconductors are
not conventional BCS systems in the sense that their normal states
deviate strongly from the predictions of Fermi liquid theory and that
the interactions are strong. Thus, a straightforward BCS self-consistent
approach should not work, particularly in view of the fact that there
isn't a well established mechanism for superconductivity in these
materials. Nevertheless, whatever actual the mechanism is, it should
describe a system with nodes and with gapless quasiparticle branches.
The actual coefficients of this effective hamiltonian cannot be derived
in a simple minded way from a microscopic system but its form will be
determined by the requirements of $d_{x^2-y^2}$ symmetry. Thus we will
use a phenomenologically-motivated BCS-like model for the quasiparticles
with nodes consistent with $d_{x^2-y^2}$ symmetry but without a self-consitent 
derivation of its
coefficients. We will consider the case of a very clean system and at
very low temperatures so as to neglect fluctuations of the amplitude of
the superconducting order parameter.

In section~\ref{sec:model} we derive an effective
Kondo-like hamiltonian for the problem of a single magnetic impurity in
an otherwise perfect $d_{x^2-y^2}$ superconductor. In this model we
focus on the effects of the quasiparticles close to the nodes of the
$d_{x^2-y^2}$ state within an energy range $\Delta_0$, the gap of the
superconductor away from the nodes. We also include, albeit in
rather crude fashion, the effects of the states above the
superconducting gap since they affect the value of the critical exchange
constant. By expanding the electron operator in the exact quasiparticle
states of the $d_{x^2-y^2}$ superconductor, we map this problem of
two-dimensional physics into an effective one-dimensional system of
chiral fermions coupled to the impurity. The effective Hamiltonian is almost
identical to the problem of a magnetic impurity in a flux phase that we
discussed in reference\cite{paperII}. The only difference here is that the
symmetry is $SU(2)$ (spin) but there are four species (or flavors) of
chiral fermions, one per node.
The mapping of the {\it electron} operator into the effective
one-dimensional fermion contains
all the information about the coherence factors of the $d_{x^2-y^2}$
state and, hence, it includes the pair breaking effects caused by the
spin fluctuations of the impurity. As a bonus, we get explicit (although
qualitative) relations between the effective coupling constants, the
relative importance of intra-node and inter-node scattering processes and
important parameters such as the location of the  impurity (relative to
the $Cu O$ plane) and the superconducting gap $\Delta_0$. We will find
that, in fact, there is always one effective channel that matters.

A considerable number of theoretical tools have been developed to study Kondo 
systems. 
For magnetic impurities in a metallic host, which have an essentially constant 
density 
of states near the Fermi energy, the different methods complement each other in 
a 
manner that we now have a rather complete understanding of this phenomenon at a 
non-perturbative level. However, with the exception of large-N methods or 
Wilson's 
numerical renormalization group, all the other methods (including the powerful 
mapping 
to one-dimensional logarithmic gases, the exact solution via the Bethe ansatz 
and the 
conformal field theory approach)  cannot be appiled to systems with a vanishing 
density 
of states. For these reasons in this work we use the large-$N_c$ approach, even 
though 
$N_c=2$ for the d-wave superconductor. In conventional Kondo systems $N_c=2$ and 
$N_c=\infty$ are known to be smoothly connected and, although it is likely that 
this 
will also hold for a d-wave system, there is still no evidence that it is also 
true for 
this problem.
In any case, given the lack of alternative approaches, we will present here a 
large-$N_c$ theory of our problem.

Our large-$N_c$ theory predicts that magnetic
impurities in {\it clean} cuprate superconductors should undergo a
quantum phase transition at a critical exchange constant  whose typical
value is crudely estimated to be below the superconducting gap
$\Delta_0$ (details are given in the next and in the last two sections).
Our estimates indicate that, for an exchange coupling $J$ with strength about 
$10 \%$ 
larger that the critical coupling,  the anomalous behaviors that we predict
should be accessible to measurements of the low temperature heat capacity and 
magnetic 
susceptibility (such as in NQR) with magnetic fields $H\sim 1-10$ Tesla and at
temperatures $T \sim 1-10 K$. The magnetic fields should be {\it in-plane}
so as not to disturb the kinematic properties of the quasiparticles on
the $CuO$ planes. At temperatures higher that $T_K$ (but still below $T_c$) the 
systems 
behaves as if it were at its quantum critical point at $J_c$. 

In ref.~\cite{paperII} we discussed the solution of the problem of a
magnetic
impurity in a flux phase problem which, as we indicated above, is closely
related the problem of a d-wave superconductor. In that work we used the
large-$N_c$ limit to investigate a similar phase transition. However, the
results that we present here for the scaling behavior of the physical
observables near this transition disagree with our previous work of
ref.~\cite{paperII}. The reason for the discrepancy is that in the process
of carrying out the large-$N_c$ limit divergent series need to be handled
and in ref.~\cite{paperII} these series were regulated in  in a manner that is
incompatible with an integer filling $Q_f$ of the impurity in the slave fermion
representation. This inconsistency is removed in this paper and the
results that we present here supersede those of ref.~\cite{paperII}. 

This paper is organized as follows. In section \ref{sec:summary} we
present a summary of the main results of this paper including experimentally
accessible predictions for magnetic impurities in high temperature
superconductors. In \ref{sec:model} we describe
the mapping of the model of a single magnetic impurity coupled to a
$d_{x^2-y^2}$
superconductor to an effective theory of chiral fermions in one-dimension
with a
non-marginal coupling.  In section \ref{sec:largeN} we discuss the
large-$N_c$ approximation
and in section \ref{sec:spe} we discuss the solution of the saddle point
equations, valid in the $N_c \to \infty$ limit, and use it to investigate
the
phase diagram of this problem at zero temperature and zero magnetic field.
In section \ref{sec:chi} we calculate the low and zero temperature
magnetic susceptibility of the impurity at zero and finite (but small)
fields in
the $N_c \to \infty$ limit. Similarly, the impurity entropy and specific
heat
(at low temperatures and fields) are calculated in section \ref{sec:ent}
again in the $N_c \to \infty$ limit. In section \ref{sec:discussion} we
conclude with a discussion of the implications of our  results and
their relation with other work, particularly the RG work of Ingersent.
Relevant details
of the computation of various integrals are given in the Appendix.

\section{Summary of Results}
\label{sec:summary}

In this section we give a brief summary of our main results and
discuss their implications for magnetic impurities in high temperature
superconductors.

\begin{enumerate}
\item
 We introduce a model for a single magnetic
impurity in a $d_{x^2-y^2}$ condensate (see section \ref{sec:model}) with
Hamiltonian $H$
\begin{equation}
H = \sum_{{\vec k}, \sigma} \epsilon({\vec k})
c^{\dagger}_{{\vec k}\sigma}
c_{{\vec k}\sigma}
- \sum_{{\vec k}} \Delta ({\vec k}) c^{\dagger}_{{\vec k}\uparrow}
c^{\dagger}_{-{\vec k}\downarrow}
+ h.c.+ {\vec S}\cdot \int d^2 x  \ J(\vec x)~c^\dagger_{\sigma}
(\vec x){\vec \tau}_{\sigma\sigma '} c_{\sigma '} (\vec x)
\label{eq:H}
\end{equation}
This is accurate at energies and
temperatures low with respect to the gap $\Delta_0$ of the $d_{x^2-y^2}$
condensate away form the nodes. We use a simple
lattice model for the $d_{x^2-y^2}$ quasiparticles (see ref.~ \cite{pines}),
which is only a cartoon of a realistic superconductor, but it has the correct
nodal structure and that is all that we actually need to know.
Next we construct an effective model of one-dimensional chiral fermions
coupled to the impurity:
\begin{eqnarray}
H_{eff} & = &
\sum_{a=1}^{4} \sum_{\sigma =\uparrow, \downarrow}
\int_{-\infty}^{\infty}
{\frac{dp}{2\pi}}~{\sqrt{vv'}}~p~d^{\dagger}_{a\sigma}(p) d_{a\sigma}(p)
\nonumber \\
& + &
\sum_{a=1}^{4} \sum_{\sigma,\nu =\uparrow, \downarrow}
(J_a /2) \left[ \int_{-\infty}^{\infty} \frac{dp}{2\pi} \sqrt{|p|}
d^{\dagger}_{a\sigma}(p)
\right] {\vec \tau}_{\sigma\nu} \cdot {\vec S}_{\rm imp}
\left[ \int_{-\infty}^{\infty} \frac{dp'}{2\pi} \sqrt{|p'|} d_{a\nu}(p')
\right]
\label{eq:1deffh-I}
\end{eqnarray}
where $v$ and $v'$ are the velocities of
the quasiparticles of the d-wave state along the two lattice directions.
This is the model that we actually investigate.
The nature of the $d_{x^2-y^2}$ state is present: 
(a) in the momentum dependence of the interaction between
the chiral fermions and the impurity (see Eq.~\ref{eq:1deffh}) and (b)
in the number and angular momenta of the channels that are coupled to
the impurity (see Eq.~\ref{eq:1dferma} and Eq.~\ref{eq:1dferm}).
We truncate the momentum dependence of the interaction beyond a momentum
scale $\sim \Delta_0/(2\pi {\sqrt{vv'}})$ where is saturates indicating
that the effective density of states is nearly constant for states above
the gap $\Delta_0$ and up to an upper cuttoff scale $D$, the bandwidth of
the normal quasiparticles $\epsilon({\vec k})$. A typical value of
$\Delta_0$ for a $CuO$ superconductor is $100 K$ and $D/\Delta_0 \geq 10$.
\item
In sections \ref{sec:largeN} and \ref{sec:spe} we solve this effective
model in the large $N_c$ limit.
The physical properties of the system in the $N_c \to \infty$
limit are determined by the behavior of the
{\it phase shift} $\delta(\epsilon)$ of Eq.~\ref{eq:finite4}.
This is the phase shift acquired by the quasiparticles of the superconductor
as they scatter off the magnetic impurity.
We find that, in contrast with the conventional Kondo problem,
the phase shift $\delta(\epsilon)$ has a strong energy
dependence. $\delta(\epsilon)$ is parametrized by the singlet amplitude
$\Delta$ and the impurity Fermi level $\epsilon_f$ (which plays the role of
the Kondo scale). These are determined by solving the saddle point equations
(Eq. \ref{eq:spe1} and Eq. \ref{eq:spe2}) which yield $\Delta$ and
$\epsilon_f$ as functions of the impurity filling $Q_f$, the exchange
constant $J$, the temperature $T$ and the magnetic field $H$.
\item
Phase Transition: We find that , at $T=H=0$, the system has a quantum
 phase transition at a
critical coupling constant $J_c \approx (2 vv'/\Delta_0)/
(1+\ln (D/\Delta_0))$. This transition separates a weak coupling free
phase in which the
impurity is nearly free with a Curie-like susceptibility, from a strong
coupling phase where the impurity is screened. We find that the
impurity is actually {\it overscreened} since the impurity magnetic
susceptibility {\it vanishes} for $J\geq J_c$ at $T=H=0$.
A na{\"\i}ve use of the BCS estimates yields $J_c \approx D \sin^2 (k_0)
/ (1+\ln (D/\Delta_0))$ which is larger than $\Delta_0$
 unless  $k_0\leq 1$ (which is not unreasonable). However, strong
coupling corrections in the superconductor (which, at best, can only be
estimated and depend strongly on the details of the mechanism of
superconductivity) in general will invalidate the simple relation between
the velocities, $k_0$ and $\Delta_0$. However, we expect that $J_c \leq
\Delta_0$, typically being of order of $\Delta_0/2$ or so. However, a
precise estimate requires a more sophisticated calculation than the one
we do here. For the puroposes of this work it will be sufficient to know
that $J_c < \Delta_0$.
\item
Kondo Scale:
We find that, close to and above  $J_c$,  the Kondo scale
$T_K=\epsilon_f$ is related to the singlet amplitude $\Delta$ by
(Eq.~\ref{eq:nu})
\begin{equation}
\epsilon_f(x,\Delta)=\Delta_0 \; {\sqrt{e}}  \; \exp \left(-{\frac{1}{\Delta}}
\;
{\frac{2x}{1-2x+\Delta}}+ {\frac{1}{1-2x+\Delta}}\right)
\left[1+O(\Delta,\Delta \ln \Delta)\right]
\label{eq:nu-I}
\end{equation}
where $e=2.7172\ldots$ and $x=Q_f/N_c$. This singular relation between
$\epsilon_f$ and $\Delta$ results from a logarithmic singularity in the
saddle point equation for the impurity occupancy. This singularity is
absent for $r <1$. In this sense, $r=1$ is like an ``upper critical
dimension" for impurity problems. The actual dependence on the coupling
constant is determined by solving the equation of state Eq.~\ref{eq:state6}
\begin{equation}
{\frac{1}{J_c}}-{\frac{1}{J}} \approx \left({\frac{\Delta_0}{\pi
vv'}}\right) \Delta + O(\Delta^3)
\label{eq:state-I}
\end{equation}
A rough estimate of $T_K$ can be obtained by setting
$Q_f=1$ and $N_c=2$. We find that, for $(J/J_c)-1 \approx
0.2$ and $\Delta_0 \approx 100 K$, $T_K=\epsilon_f \approx 9 K$, while , for
$(J/J_c)-1 \approx 0.1$, $T_K \approx 1 K$.
\item
We have calculated the impurity susceptibility and specific heat in the
overscreened phase $J>J_c$ for magnetic fields and temperatures $H,T < T_K$
which can be realized unless $J$ is too close to $J_c$. For $T \ll H \ll T_K$
we find (Eq.~\ref{eq:chi9}) the susceptibility
$\chi_{\rm imp}~\sim~
N_c \left(\frac{\Delta}{\epsilon_f}\right)^2~H\ln
\frac{\Delta_0}{H}$
while , in the opposite regime $H \ll T \ll T_K$ (Eq.~\ref{eq:chilow})
$\chi_{\rm imp}~\sim~ 2 N_c \ln 2  \;
\left({\frac{\Delta}{\epsilon_f}}\right)^2 T \ln
({\frac{\Delta_0}{T}})$.
Similarly, the specific heat in the regime $H \ll T \ll T_K$ is
(Eq.~\ref{eq:Clow})
$C_{\rm imp}(0,T) \approx 9 \zeta(3) N_c
{\frac{\Delta^2}{\epsilon_f^2}} T^2 \ln({\frac{\Delta_0}{T}})$, while
for $T \ll H \ll T_K$ we find instead the result (Eq.~\ref{eq:Chigh1})
$C_{\rm imp}(H,T) \approx  N_c {\frac{\pi^2}{3}}   \;\;
\left({\frac{\Delta}{\epsilon_f}}\right)^2 T~ H \ln
({\frac{\Delta_0}{H}})$. The low field regime is clearly very different
from a Fermi liquid although a Wilson Ratio can still be defined and it is
finite (Eq.~\ref{eq:Wlow})
${\frac{C_{\rm imp}(H,T)}{T \chi_{\rm imp}(H,T)}}
\approx {\frac{9 \zeta(3)}{2 \ln 2}}$. The behavior in the
high field regime is more like a Fermi liquid.
These behaviors should be accessible to experiments
in clean samples of cuprate superconductors
at magnetic fields of $1-10$ Tesla.

We have not investigated yet the quantum critical regime
$H,T > T_K$, which we will discuss in a separate publication\cite{paper4}.
\end{enumerate}

\section{The Model}
\label{sec:model}

In this section we construct the model that describes the coupling of the
quasi-particles of a {\it d-wave} superconductor to a localized magnetic
impurity. We will show explicitly that this model maps exactly onto a model
of a magnetic impurity coupled to the spinons of a flux phase that we
discussed in ref.~\cite{paperII}. The strategy that we will follow in this
section consists of first writing down a simple model for the quasi-particles
of the d-wave superconductor with a physically reasonable coupling to a local
magnetic moment. Next we will carry a dimensional reduction of this problem
down to a model of effective one-dimensional (chiral) fermions which has all
the symmetries of the d-wave superconductor. The effective one-dimensional
model coincides exactly with the non-marginal Kondo problem that was discussed
in reference \cite{paperII}. In the next section we will use the results
of \cite{paperII} to draw conclusions on the effects of magnetic impurities
in d-wave superconductors.

\subsection{Free Hamiltonian}
\label{subsec:free}

We begin by choosing a model of a d-wave superconductor with the form of
a
BCS-type Hamiltonian. It has a kinetic energy term (which we choose to be
of
the form of a tight-binding Hamiltonian) and a pairing term with d-wave
symmetry. In subsection~\ref{subsec:imp} we will describe the way magnetic
impurities couple to the quasi-particles. Here we will make the
phenomenological
assumption that there is d-wave pairing regardless of the mechanism that
gives
rise to that pairing. BCS-type models which exhibit d-wave pairing (driven by
antiferromagnetic fluctuations) have been proposed by Bickers, Scalapino
and
White~\cite{doug} and by Monthoux and Pines~\cite{pines}.
Here we will use a BCS model of this type to describe the dynamics of the
quasi-particles.

What will be important for the dynamics is that the model exhibits four
 nodes
where the gap vanishes and that the gap is fairly large away from the
nodes.
Thus, we will concentrate on the behavior of the quasi-particles close to
the
nodes. Instead of using the full detailed form of the gap, we will replace it
by
a linearized spectrum with a wavevector cutoff $\Lambda$ (relative to the
location of the node) such that the energy of the quasi-particles with
wavevector $\Lambda$ is approximately equal to the value $\Delta_0$ of the
superconductor gap away from the nodes. The actual structure of the
quasi-particle spectrum away from the nodes will play very little role and such
states will be neglected. This view is supported  by recent
photoemmission
experiments by Shen {\it et.al.,\/}\cite{shen}~ and Ding
{\it et.al.,\/}\cite{campuzano}~in $YBaCuO$ superconductors where a Fermi
surface is seen at optimal doping and it disappears progressively away
from
the nodes (where the gap vanishes) for underdoped systems.
Hence, the important features of the quasi-particle spectrum that we
will keep
are the four nodes where the excitations are gapless, the correct behavior
under
lattice symmetries (and parity) and the Fermi velocities at the nodes.

The Hamiltonian for the quasi-particles of a BCS-type superconductor
in the
absence of impurities is
\begin{equation}
H_0 = \sum_{{\vec k}, \sigma} \epsilon({\vec k})
c^{\dagger}_{{\vec k}\sigma}
c_{{\vec k}\sigma}
- \sum_{{\vec k}} \Delta ({\vec k}) c^{\dagger}_{{\vec k}\uparrow}
c^{\dagger}_{-{\vec k}\downarrow}
+ h.c.
\end{equation}
To make the model concrete we use a lattice model for the
quasi-particles with
a
bare energy $\epsilon({\vec k})$ of the form
\begin{equation}
\epsilon({\vec k}) = \epsilon(-{\vec k}) = -2 t \left( \cos(k_1) +
\cos(k_2)
\right) + \mu
\end{equation}
The Fermi operators $ c^{\dagger}_{{\vec k}\sigma}$
create quasi-particles with
momentum $\vec k$, spin $\sigma$ and energy $\epsilon({\vec k})$.
Here $\mu$ is
the chemical potential for the quasi-particles.
The gap function $\Delta ({\vec k})$ for a $d_{x^2-y^2}$ superconductor
given by
\begin{equation}
\Delta(k) = \Delta_0 \left(\cos(k_1) - \cos(k_2)\right)
\end{equation}
Here $\Delta_0$ set the scale for the gap away from the nodes.
The one-particle
spectrum of this simple Hamiltonian agrees qualitatively with the
observed
photoemission spectrum.

Next we write the quasi-particle operators in the Nambu-Gorkov
form~\cite{schrieffer}
\begin{equation}
\Phi({\vec k}) = \left( \matrix{ c_{{\vec k}\uparrow}
\cr c^\dagger_{-{\vec k}\downarrow}} \right)
\label{eq:ng}
\end{equation}
In terms of the Nambu-Gorkov spinors, the free part of the Hamiltonian
$H_0$ can now be written as
\begin{equation}
H_0=\sum_{{\vec k}}^{} \Phi^{\dagger}({\vec k})
\left[ \left( \epsilon({\vec k}) -
\mu\right)\tau_3
- \Delta({\vec k}) \tau_1\right] \Phi({\vec k})
\end{equation}
where $\tau_1$ and $\tau_3$ are two Pauli matrices.
This model has four  nodes~\cite{wenger} at the points in the
Brillouin
zone given by $\left( \pm k_0, \pm k_0\right)$, with $k_0 \equiv
\arccos(\mu/4t)$.
The spectrum of the quasi-particles crosses the Fermi surface at those
points
and the gap closes. As a consequence, the quasi-particles have a linear
dispersion relation in the vicinity of these {\it nodes}. This can be
shown~\cite{wenger} by expanding for small momentum departures around
the $\left(\pm k_0, \pm k_0\right)$ points. 

The next step in the construction of an effective low-energy model is to
describe the dynamics of the quasi-particles close to the nodes. To this
end we
assign a label for each one of the four nodes. Let $a=1,2,3,4$ be this
label
and we assign the label $a=1$ to the node $(k_0,k_0)$, $a=2$ to the node
$(-k_0,-k_0)$, $a=3$ to the node $(-k_0,k_0)$ and $a=4$ to the node
$(k_0,-k_0)$. Let $\vec q$ be the momentum relative to the node. It is useful to 
work in the rotated basis 
$p_1 \equiv (1/\sqrt{2})(q_1 + q_2)$ and $p_2 \equiv (1/\sqrt{2})
(q_1 - q_2)$,
with velocities
$ v \equiv 2\sqrt{2} t \sin(k_0)$,
$ v' \equiv \sqrt{2}\Delta_0 \sin(k_0)$, where $\Delta_0$ is the size of
the
superconductor gap at its maximum value. Let $\Phi_a^{\dagger}(\vec p)$
denote
the (Nambu-Gorkov spinor) operator which creates a quasi-particle with
(rotated)
momentum $\vec p$ {\it relative} to the wavevector of node $a$. The free
Hamiltonian now takes the form
\begin{eqnarray}
H_0 & = &
\int \frac{d^2 p}{(2\pi)^2}
\left\{
\Phi_{1}^{\dagger} (\vec p) \left( v p_1 \tau_3 + v' p_2 \tau_1
\right) \Phi_{1} (\vec p)
-
\Phi_{2}^{\dagger} (\vec p) \left( v p_1 \tau_3 + v' p_2 \tau_1
\right) \Phi_{2} (\vec p)
\right\}
\nonumber \\
& - &
\int \frac{d^2 p}{(2\pi)^2}
\left\{
\Phi_{3}^{\dagger} (\vec p) \left( v p_2 \tau_3 + v' p_1 \tau_1
\right) \Phi_{3} (\vec p)
-
\Phi_{4}^{\dagger} (\vec p) \left( v p_1 \tau_3 + v' p_2 \tau_1
\right) \Phi_{4} (\vec p)
\right\}
\label{eq:kinen0}
\end{eqnarray}
In the long-wavelength limit the Hamiltonian splits into four
(anisotropic)
Dirac-like hamiltonians. In what follows we will refer to these
four sets of
excitations (which represent the four nodes of the d-wave
superconductor) as to
the four {\it flavors} (or channels).

It will prove useful for our purposes to rotate
$\Phi_a(\vec k)$ to a new field $\psi_a$
\begin{equation}
\Phi_a(\vec k)={\frac{1}{\sqrt{2}}} (1-i \tau_1)\psi_a(\vec k)
\label{eq:fields}
\end{equation}
and to write $H_0$ in terms of $\psi_a$,
\begin{eqnarray}
H_0 & = & \int \frac{d^2p}{(2\pi)^2}
\left(
\matrix{ 0 & \epsilon_+ e^{-i\theta_+}\cr \epsilon_+ e^{i\theta_+} & 0\cr}
\right)
( \psi^{\dagger}_1(\vec p) \psi_1(\vec p)-\psi^{\dagger}_2(\vec p)
\psi_2(\vec
p))
\nonumber \\
& & \qquad\quad -
\int \frac{dp^2}{(2\pi)^2}
 \left(
\matrix{ 0 & \epsilon_- e^{-i\theta_-}\cr \epsilon_- e^{i\theta_-} & 0\cr}
\right)
( \psi^{\dagger}_3(\vec p) \psi_3(\vec p)-\psi^{\dagger}_4(\vec p)
\psi_4(\vec
p))
\label{eq:freeh}
\end{eqnarray}
In Eq.~(\ref{eq:freeh})
the following definitions have been used:
\[ \epsilon_+ \equiv \sqrt{(vp_1)^2 + (v'p_2)^2}
\equiv \sqrt{v v'} p_+ ;\quad
\theta_+ \equiv \tan^{-1}\left( vp_1/v'p_2\right);\]
\[
\epsilon_- \equiv \sqrt{(v'p_1)^2 + (vp_2)^2}
\equiv \sqrt{v v'} p_- ; \quad
\theta_- \equiv \tan^{-1}\left( vp_2/v'p_1\right)
\]
where $ v \equiv 2\sqrt{2} t \sin(k_0)$, \quad
$ v \equiv \sqrt{2}\Delta_0 \sin(k_0)$.

Next we notice the fact that, as far as the kinetic energy is concerned,
$p_+$, $p_-$, $\theta_+$, $\theta_-$ are just dummy variables and that
the
measure in the integrals is invariant under the change $p_1, p_2$ into
$p_+,
p_-$. It turns out, an we show this below, that the interaction term is
also
invariant under a redefinition of the integration variables.
Naturally, the quasi-particle operators themselves are not invariant
under
these
redefinitions of variables. Hence, although all explicit reference to
the
anisotropy can be removed from the Hamiltonian, it remains quite explicit
 in
the relation between the quasi-particle (fermion) operators and the
fields that
will describe the effective Hamiltonian, {\it i.e.,\/} in generalized
coherence factors.

Taking these observations into consideration, $H_0$ can be put in a much
simpler form
\begin{equation}
H_0  =
\int_0^{\infty} \frac{p~dp}{2\pi}
\int_0^{2\pi} \frac{d\theta}{2\pi}
\left(
\matrix{ 0 & \sqrt{vv'} p e^{-i\theta}\cr \sqrt{vv'} p e^{i\theta} & 0\cr}
\right)
\sum_{a=1}^4
 \psi^{\dagger}_a(\vec p) T_{ab} \psi_b(\vec p)
\label{freeh2}
\end{equation}
where $T_{ab}$ is the $4 \times 4$ diagonal matrix in flavor indices
${\rm diag}(1,-1,-1,1)$. The signs in the matrix $T_{ab}$ account for the
{\underline {parity}} of each node.
Here we have only kept explicitly the flavor (node) indices.

We now diagonalize the kinetic energy and expand the fields in energy
eigenmodes.
\begin{equation}
\psi(\vec p)  = \psi_+(\vec p)  u_+(\theta) + \psi_-(\vec p) u_-(\theta)
\label{eq:exppsi1}
\end{equation}
where
\begin{equation}
u_{\pm} (\theta) =  \frac{1}{\sqrt{2}}
\left(
\begin{array}{l}
1 \\ \pm e^{i\theta}
\end{array}
\right)
\label{eq:eigenvectors}
\end{equation}
are the spinors that diagonalize the d-wave BCS Hamiltonian near the
nodes. The effective rotational invariance around each node
(in terms of the redefined momenta) enables us to expand in angular momentum 
eigenmodes 
around each node
\begin{equation}
\psi_\pm (\vec p) = \sum_{m=-\infty}^{\infty} e^{i m \theta}
\psi_{\pm m} (|p|)
\label{eq:exppsi2}
\end{equation}
This is effectively an
angular momentum expansion in elliptic coordinates around each node.

$H_0$ is now diagonal and takes the simpler form
\begin{equation}
H_0  =   \sum_{a=1}^{4} \int_0^{\infty} \frac{p~dp}{2\pi}
\sqrt{vv'} p
\sum_{m=-\infty}^{\infty} T_{ab}\left[
\psi^{\dagger}_{a,+,m} (|p|) \psi_{b,+,m}(|p|)
- \psi^{\dagger}_{a,-,m}(|p|) \psi_{b,-,m}(|p|)\right]
\label{kinen}
\end{equation}

\subsection{Impurity Interaction}
\label{subsec:imp}

Now we consider the interaction term for spin impurities given by
\begin{equation}
H_{\rm imp} \equiv {\vec S}\cdot \int d^2 x  \ J(\vec x)~c^\dagger_{\sigma}
(\vec x){\vec \tau}_{\sigma\sigma '} c_{\sigma '} (\vec x)
\label{eq:imp1}
\end{equation}
In practice we will be interested in well localized impurities. This means that
$J(\vec x)$ is sharply peaked at some point ${\vec x}_0$ where the
impurity is
located.
Realistic magnetic impurities in $YBaCuO$ and other High Temperature
Superconductors~\cite{pines} almost always involve magnetic atoms which
either substitute a $Cu$ atom or hybridize strongly with it.
This is the case for $Ni$ which, due to its hybridization with oxygen,
it is believed to behave like a $S=1/2$ impurity spin~\cite{pines}.
Similarly, $Zn$ substitutes
$Cu$ which now behaves like a missing $S=1/2$ magnetic moment and in this
sense is a magnetic impurity. In all cases of $Cu$ substitution we will
model
the impurity as a localized $S=1/2$ moment residing at a site of the
square
lattice which we will consider as the origin. Notice, however, that $O$
can also behave like a magnetic impurity in the cuprates.  An $O$
magnetic
impurity sits in the middle of the bond instead of a corner $Cu$ site.
This
case leads to more complicated form of the effective interaction which
we will
not discuss in this thesis.

The effects of magnetic impurities on $Cu$ sites can be modeled
qualitatively
in terms of an exchange coupling constant $J(\vec x)$ which couples most
strongly to the quasi-particles at ${\vec x}=0$ and decays rapidly and
symmetrically around ${\vec x}=0$. For simplicity we will use a model
in which
$J(\vec x)$ is a narrow gaussian.
We can see clearly from the discussion that led to
the effective free Hamiltonian, that the only properties of $J(\vec x)$
that
are important are the amplitudes of its Fourier transform at the relative
wavevector of the nodes. These amplitudes play the role of the effective
coupling constants. Physically, the strength of the exchange coupling is
determined by an overlap integral which decays very quickly. Thus,
impurities
which substitute ${\rm Cu}$ atoms {\it in the plane} are more strongly
coupled
than those that substitute ${\rm Cu}$ {\it out of the plane}. Also
impurities
on sites
other than $Cu$ sites are more weakly coupled to the quasi-particles
than those
on $Cu$ sites. These observations are important since we will see in
section~\ref{sec:spe} that the impurities are Kondo screened if their
exchange coupling constants are large enough.

We now proceed to find the contribution of the impurity interaction to the
effective Hamiltonian. In momentum space Eq.~(\ref{eq:imp1}) becomes
\begin{equation}
H_{\rm imp} \equiv
\int \frac{d^2k}{(2\pi)^2}  \int \frac{d^2k'}{(2\pi)^2}
J({\vec k}-{\vec k}')~{\vec S}\cdot~c^{\dagger}_\sigma(\vec k)
{\vec \tau}_{\sigma\sigma'}
c_{\sigma'}({\vec k}')
\label{imph0}
\end{equation}
In terms of the  NG spinors it reads
\begin{equation}
H_{\rm imp} =
\int_{{\vec k}, {\vec k}'}  \sum_{i, j=1}^{2}
\left\{
J({\vec k}-{\vec k}') \left[
{S}_3~\Phi_i^{\dagger} ({\vec k})\Phi_i({\vec k}')
+  {S}_-~\epsilon_{ij} \Phi^{\dagger}_i ({\vec k})\Phi^{\dagger}_j
(-{\vec k}')
+ h.c.
\right]
\right\}
\label{imph1}
\end{equation}
where $S_j$ represents the impurity spin, $S_- \equiv \frac{1}{2}
( S_x - i~S_y )$ and $\epsilon_{ij}$ is a $2\times 2$ skew symmetric
tensor.

As before, we expand the NG spinors in their components centered around
the nodes. Since we have four nodes  the impurity Hamiltonian has terms
which
describe spin flip scattering processes involving, in addition, eventual
inter-node
scattering processes. The strength of these scattering processes is
determined
by $J(\vec Q)$ where $\vec Q$ is the relative wavevector of a pair of
nodes.
There are four cases of interest:
\begin{enumerate}
\item
${\vec Q} \sim 0$, corresponding to scattering processes that do not
mix nodes
(``forward scattering").
The corresponding coupling constant is $J(0)\equiv J_0$.
\item
${\vec Q} \sim 2k_0 {\hat e}_1$, which mixes nodes $1$ with $3$ and $2$
with
$4$. This coupling constant is $J_1$.
\item
${\vec Q} \sim 2k_0 {\hat e}_2$, which mixes nodes $1$ with $4$ and $2$
with
$3$. This coupling constant is $J_2$. For systems with exact tetragonal
(square) symmetry $J_1=J_2$.
\item
${\vec Q} \sim 2k_0 ({\hat e}_1 \pm {\hat e}_2$, which mixes nodes $1$
with $2$
and $3$ with $4$. These coupling constants are $J_d^\pm$. For tetragonal
systems they reduce to just one (diagonal) coupling $J_d$.
\end{enumerate}
For example, consider an impurity seated at the Cu site at $x=0$.
As a crude approximation we may assume $J({\vec x}) \approx {\bar J}
\delta
({\vec x})$.
The Fourier transform  tells us that all the couplings will be the same,
and
equal to $\bar J$.
A more realistic shape for $J({\vec x})$ would be a gaussian centered at
the impurity site ${\vec x}=0$ and decaying rapidly within a distance of
the
order of a  lattice constant $\lambda$.
Thus  we take $J(x) \approx {\bar J}/(2\pi \lambda^2)~e^{-(1/2\lambda^2)
{\vec x}^2}$ will generate, for a generic $k$-vector (which in our
case will be~$2k_0 {\hat e}_1$,~$2k_0{\hat e}_2$,~$2k_0 ( {\hat e}_1
+ {\hat e}_2)$~and~$2k_0({\hat e}_1 - {\hat e}_2)$),
\begin{equation}
J(\vec k)=
{\bar J}~e^{-{\frac{1}{2}}\lambda^2 k^2}
\label{jotafourier}
\end{equation}
For the impurity at the origin in a tetragonal (square) lattice,  all the
coupling
constants are real, with $J_0~>~J_1 = J_2~>~J_d$.
In the language of the fields introduced in Eq.~(\ref{eq:fields})
the impurity
Hamiltonian now becomes
\begin{eqnarray}
H_{\rm imp} & = &
S_3~\sum_{a,b=1}^{4}{\rm K}^3_{\ a b}~ \sum_{i=1}^{2}
\int \frac{d^2p}{(2\pi)^2}
\psi_{a,i}^{\dagger}(\vec p)
~\int \frac{d^2 p'}{(2\pi)^2} \psi_{b,i} ({\vec p}') \nonumber \\
& + &
S_-~\sum_{a,b=1}^{4}{\rm K}^+_{\ a b}~ \sum_{i=1}^{2} ~
\int \frac{d^2p}{(2\pi)^2} \psi^{\dagger}_{a,i} (\vec p)~(i\tau_2)_{i,j}~
\int \frac{d^2 p'}{(2\pi)^2}\psi^{\dagger}_{b,j}(-p')
+ {\rm h~.c.~}
\label{eq:imph2}
\end{eqnarray}
In Eq.~(\ref{eq:imph2}) the indices $a, b$ are the flavor indices which
label
the
effective Dirac fermions species associated with each node.
The indices $i,j$ run
through the spinor components (two per each NG spinor, {\it i.e.,\/}
per node)
and label linear combinations of quasi-particles with spin up with holes
with
spin down.
Also notice that $p$ and $p'$ now label small departures
from the appropriate node.
Using the fact that $i \tau_2$ is an antisymmetric matrix, we can rewrite
Eqn.(\ref{eq:imph2}) in the form
\begin{eqnarray}
H_{\rm imp} & = &
S_3~\sum_{a,b=1}^{4}{\rm K}^3_{\ a b}~ \sum_{i=1}^{2}
\int_0^{\infty} p\frac{dp}{2\pi} \int_0^{2\pi} \frac{d\theta}{2\pi}
\psi_{a,i}^{\dagger}(\vec p)
\int_0^{\infty}p' \frac{d p'}{2\pi} \int_0^{2\pi} \frac{d\theta'}{2\pi}
\psi_{b,i} ({\vec p}') \nonumber \\
& + &
S_-~\sum_{a,b=1}^{4}{\rm K}^+_{\ a b}~ \sum_{i=1}^{2} ~
\int_0^{\infty} p\frac{dp}{2\pi} \int_0^{2\pi} \frac{d\theta}{2\pi}
\psi^{\dagger}_{a, 1} (\vec p)
\int_0^{\infty} p' \frac{d p'}{2\pi} \int_0^{2\pi} \frac{d\theta'}{2\pi}
\psi^{\dagger}_{b, 2}(-p')
+ {\rm h.~c.~}
\label{eq:imph3}
\end{eqnarray}
The $4\times 4$ matrices ${\rm K}^3_{ab}$ and
${\rm K}^+_{ab}$ used in Eqns.~(\ref{eq:imph2}) and (\ref{eq:imph3})
are given
by
\begin{eqnarray}
{\rm K}^3_{ab} =
\left(
\begin{array}{llll}
J_0  &  J_d & J_1 & J_2 \\
J_d  & J_0  & J_2  & J_1 \\
J_1 & J_2 & J_0 & J_d \\
J_2 & J_1 & J_d & J_0
\end{array}
\right);
\quad
{\rm K}^+_{ab} =
\left(
\begin{array}{llll}
J_d & J_0 & J_2 & J_1 \\
J_0 & J_d & J_1 & J_2 \\
J_2 & J_1 & J_d & J_0 \\
J_1 & J_2 & J_0 & J_d
\end{array}
\right)
\end{eqnarray}
The form of Eqn.(\ref{eq:imph3}) strongly suggests the
following change of variables (particle-hole transformations) performed on the
second component
of all four flavors
\begin{equation}
\psi_{1,2} (p) \rightarrow \psi^{\dagger}_{1,2} (-p);
\ \
\psi_{2,2} (p) \rightarrow \psi^{\dagger}_{2,2} (-p);
\ \
\psi_{3,2} (p) \rightarrow \psi^{\dagger}_{3,2} (-p);
\ \
\psi_{4,2} (p) \rightarrow \psi^{\dagger}_{4,2} (-p)
\label{eq:ptclehole}
\end{equation}
to express the interaction term as a scalar product of two spin-$\frac{1}{2}$
operators. 

We can now separate the modes and find an effective one-dimensional model.
After integration over the angle variable $\theta$, the fields involved in
Eqn.(\ref{eq:imph3}) become
\begin{eqnarray}
\int_0^{2\pi} \frac{d\theta}{2\pi} \psi_{1 a}
& = & \frac{1}{\sqrt{2}}
\left[ \psi_{0+}(|p|) + \psi_{0-}(|p|)\right]_{a}
\nonumber \\
\int_0^{2\pi} \frac{d\theta}{2\pi} \psi_{2 a}
& = & \frac{1}{\sqrt{2}}
\left[ \psi_{-1 +}(|p|) - \psi_{-1-}(|p|)\right]_{a}
\label{components}
\end{eqnarray}
Now we define, for each flavor $a$, an effective one-dimensional
{\it chiral}
(right moving) fermi field
\begin{eqnarray}
d_{1 a} (p) & \equiv &
\left\{
\begin{array}{l}
\sqrt{|p|}~\psi_{0,+,a} (|p|); \quad {\rm for} \quad p>0;
\\
\sqrt{|p|}~\psi_{0,-,a} (|p|); \quad {\rm for} \quad p<0;
\end{array}
\right.
\label{eq:1dferma}
\\
d_{2 a} (p) & \equiv &
\left\{
\begin{array}{l}
\sqrt{|p|}~\psi_{-1,+,a} (|p|); \quad {\rm for} \quad p>0;
\\
-~\sqrt{|p|}~\psi_{-1,-,a} (|p|); \quad {\rm for} \quad p<0;
\end{array}
\right.
\label{eq:1dferm}
\end{eqnarray}
and Eqn.(\ref{eq:imph3}) can be recast as
\begin{eqnarray}
H_{\rm imp}
& = &
S_3~\frac{1}{2} \sum_{ab,i}^{} K_{ab}^3
\int_{-\infty}^{\infty} \frac{dp}{2\pi} \sqrt{|p|}~
d^{\dagger}_{ai} (p)
\int_{-\infty}^{\infty} \frac{dp'}{2\pi} \sqrt{|p'|}~
d_{bi} (p')
\nonumber \\
& + &
S_-~\sum_{ab}^{} K_{ab}^+
\int_{-\infty}^{\infty} \frac{dp}{2\pi} \sqrt{|p|}~
d^{\dagger}_{a 1} (p)
\int_{-\infty}^{\infty} \frac{dp'}{2\pi} \sqrt{|p'|}~
d^{\dagger}_{b2} (p')
+ h.c.
\label{eq:imph4}
\end{eqnarray}
Here we perform the change of variables suggested above by setting
\begin{equation}
d_{a 2} (p) \to d_{a 2}^{\dagger} (-p), \quad {\rm for} \quad a=1,..,4
\label{eq:phole}
\end{equation}
and thus, in the definition given by Eqn.(\ref{eq:1dferm})
we rename  $d_{1a}(p)$ as the $d_{\uparrow a}(p)$ component of an
effective spin-$\frac{1}{2}$ one-dimensional chiral fermion,
and $d_{2 a}^{\dagger} (-p)$ as the $d_{\downarrow a}(p)$ component.
Please
notice that this label is {\it not} equivalent to the spin of the
original
quasi-particles. In fact, the relation between these effective
one-dimensional
chiral fermions and the original quasi-particles is actually quite
complicated.
The (flavor) coupling matrices commute with each other (as required
by the
$SU(2)$ spin rotation invariance) and can be diagonalized simultaneously
by
means of the following unitary transformation
\begin{equation}
d'_{a i} = {\rm U}_{ab}~d_{bi}
\label{eq:flavrot1}
\end{equation}
where $i = \uparrow {\rm or} \downarrow$, the flavor indices
$a$ and $b$ run from 1 to 4 and
\begin{equation}
{\rm U}_{ab} = \frac{1}{2}
\left(
\begin{array}{rrrr}
1 & 1 & 1&1\\
-1&1&-1&1\\
-1&-1&1&1\\
1&-1&-1&1
\end{array}
\right)
\label{flavrot2}
\end{equation}
This rotation brings the coupling matrices ${\rm K}^3_{ab}$
and ${\rm K}^+_{ab}$ to the diagonal form
\begin{eqnarray}
{\rm K}^3_{ab} =
\left(
\begin{array}{cccc}
J'_1&0&0&0
\\
0&J'_2&0&0
\\
0&0&J'_3&0
\\
0&0&0&J'_4
\end{array}
\right);
\quad
{\rm K}^+_{ab} =
\left(
\begin{array}{cccc}
J'_1&0&0&0
\\
0&-J'_2&0&0
\\
0&0&J'_3&0
\\
0&0&0&-J'_4
\end{array}
\right);
\label{eq:K'}
\end{eqnarray}
with
\begin{equation}
\begin{array}{l}
J'_1  =  J_0 + J_d + J_1 + J_2
\\
J'_2  =  J_0 - J_d + J_1 - J_2
\\
J'_3  =  J_0 + J_d - J_1 - J_2
\\
J'_4  =  J_0 - J_d - J_1 + J_2
\end{array}
\label{eq:couplings}
\end{equation}
As one can see in Eqn.(\ref{eq:K'}) flavors 2 and 4 appear to
have $S_x$ and $S_y$ with the sign reversed.
However this can be compensated by the following additional rotation
in the
spin components
\begin{eqnarray}
d_{2\uparrow}(p) & \to & i~d_{2\uparrow}(p),
\qquad
d_{2\downarrow}(p) \to -i~d_{2\downarrow}(p);
\nonumber \\
d_{4\uparrow}(p) & \to & i~d_{4\uparrow}(p),
\qquad
d_{4\downarrow}(p) \to -i~d_{4\downarrow}(p)
\label{eq:spinrot}
\end{eqnarray}
After all of these manipulations we find that the effective
one-dimensional theory for this model is
\begin{eqnarray}
H_{eff} & = &
\sum_{a=1}^{4} \sum_{\sigma =\uparrow, \downarrow}
\int_{-\infty}^{\infty}
\frac{dp}{2\pi} E(p) d^{\dagger}_{a\sigma}(p) d_{a\sigma}(p)
\nonumber \\
& + &
\sum_{a=1}^{4} \sum_{\sigma,\nu =\uparrow, \downarrow}
(J_a /2) \left[ \int_{-\infty}^{\infty} \frac{dp}{2\pi} \sqrt{|p|}
d^{\dagger}_{a\sigma}(p)
\right] {\vec \tau}_{\sigma\nu} \cdot {\vec S}_{\rm imp}
\left[ \int_{-\infty}^{\infty} \frac{dp'}{2\pi} \sqrt{|p'|} d_{a\nu}(p')
\right]
\label{eq:1deffh}
\end{eqnarray}
In Eqn.(\ref{eq:1deffh}) we dropped the primes in Eqn.(\ref{eq:flavrot1})
and in the effective coupling constants. The kinetic energy of the chiral
fermions is $E(p)={\sqrt{vv'}} p$.

Eq.~(\ref{eq:1deffh}) can be recognized to be {\underline {exactly}}
the non-marginal Kondo
Hamiltonian that was discussed in ref.~\cite{paperII}. Hence, the
effective
Hamiltonian for a d-wave superconductor coupled to a magnetic
impurity is
essentially  equivalent to a (multichannel) generalization of a
non-marginal
Kondo problem. There are four channels, one for each node. The channel
degeneracy is generally lifted by the inter-node scattering. In fact,
Eq.~(\ref{eq:couplings}) shows that in the absence of inter-node
scattering
({\it i.e.,\/} $J_1=J_2=J_d=0$) the four flavors couple to the
impurity with
exactly the same exchange interaction strength $J'_a=J_0$
($a=1, \ldots, 4$).
For a strictly tetragonal system the couplings are ordered in the
sequence
$J_1'>J_2'=J_4'>J_3'$. Intuitively one expects the channel with the
largest
coupling to dominate the low energy limit. In the extreme limit in
which all
inter-node and intra-node amplitudes are exactly equal one finds that
channels
2, 3 and 4 decouple and that only the remaining channel 1 couples to the
impurity. Thus, in this limit, the physics of the system is that of a
single
channel non-marginal Kondo problem.

Given that these two seemingly different systems are actually equivalent,
most
of the results found in ref.~\cite{paperII} carry over to this problem
almost
without change but with a new physical meaning and processes,
in particular including pair breaking effects. In ~\cite{paperII} we
found that there is a critical value of the exchange coupling constant
$J_c$,
above which the impurity spin is screened. We also found there that the
critical value $J_c$ was of the same order as the energy cutoff, which
here is the superconducting gap $\Delta_0$. The reason behind the
existence of a finite $J_c$ is that the effective interaction between the
impurity
and the normal excitations is momentum dependent and that it vanishes at
small momenta (see Eq.~\ref{eq:1deffh}). However, the same momentum
dependence
makes the effective coupling  grow arbitrarily large at large momenta.
This last behavior is unphysical and it results from the approximations,
 which are accurate at small momenta only. This observation motivates
a simple redefinition of the model with a  finite, momentum independent,
coupling at momenta larger than a scale of the order of
$\Delta_0/(2\pi{\sqrt{v v'}})$.

In ref.~\cite{paperII} we showed that a momentum-dependent coupling is
equivalent
to a change in the density of states (DOS) for a theory with a momentum
independent coupling constant. The model of ref.~\cite{paperII},
and the model
discussed above, have a DOS vanishing linearly with the energy.
We consider
a modified model with the DOS
\begin{equation}
\rho(\epsilon) = \left\{
\begin{array}{l}
{\frac{|\epsilon|}{2\pi vv'}}
\qquad {\rm for} \qquad |\epsilon| \leq \Delta_0 \cr
{\frac{\Delta_0 }{2\pi vv'}}\qquad {\rm for} \qquad
\Delta_0 < \ |\epsilon| \ < D
\end{array}
\right.
\label{form1}
\end{equation}
where $\Delta_0$ is the size of the superconductor gap away from
the nodes.
This change in the DOS is equivalent to a saturation of the coupling
constant at the momentum scale $\Delta_0/(2\pi{\sqrt{v v'}})$.

In other words, we are assuming a linear dependence of the DOS with the
energy
around the gap nodes, up to the energy scale of the superconductor gap.
For energies higher than the superconducting gap $\Delta_0$, the normal
quasiparticles are, for all practical purposes, identical to normal
electrons.
In a realistic cuprate superconductor, the band structure is actually
rather
complicated. Nevetheless, we can take into account the contribution of
these
states to the physics by considering a flat fermion band characteristic
of a
continuum spectrum from $\Delta_0$ up to a bandwidth $D$,
which works as a high energy cutoff. As we will see below, the
contribution of these states can almost always be ignored but they will
enter in our results in two important places: (a) by shifting
(downwards) the critical value of the coupling constant $J_c$ and (b)
in the
scaling behavior for ``half-filled" impurities.
The shift in $J_c$ is quantitatively important and it results in a
downwards shift of $J_c$ from the nominal value of the superconducting
gap $\Delta_0$. Hence, we will assume that $J_c < \Delta_0$.
This happens if the scales of $\Delta_0$ and $D$ are reasonably well
separated.

\section{Large $N_c$ Theory}
\label{sec:largeN}

In the previous section we constructed a model for a magnetic impurity
embedded in a d-wave superconductor and showed that it is equivalent to
a special non-marginal Kondo problem. In this section we solve this
model in the large $N_c$ approximation, where $N_c$ is the rank of the
symmetry group of the impurity spin. In the physically relevant
situation $N_c=2$ ({\it i.~e.~\/}, spin one-half). Clearly, in this
situation $N_c$ is not large. Nevertheless we expect the large $N_c$
theory to give a qualitatively correct description. We now proceed with
a brief summary of the large-$N_c$ theory\cite{read} as
adapted\cite{paperII}
to the physical situation described by the Hamiltonians of the previous
section.

In order explore the physics of this system we extend the symmetry
from $SU(2)$ (spin) to $SU(N_c)$ and look at it within the
large-$N_c$ approximation. Notice that, unlike the Coqblin-Schrieffer
model, $N_c$ is not related to a magnetic impurity in a
higher spin representation. Similarly, the four flavors of fermions
originate  from the nodal structure of the superconductor and are not
related to an orbital-degeneracy as in the multichannel Kondo problem in
metals. Thus the problem we want to study has $N_c=2$ ``colors". The
number of ``flavors"  is $N_f=1$ if there is node mixing and
$N_f=4$ in the abscence of inter-node scattering. However, there is a
subtlety in the treatment of the impurity once
symmetry is extended from $SU(2)$ to $SU(N_c)$. For the group $SU(2)$,
the lowest representation for an impurity is $S=1/2$ . For $SU(N_c)$
many more representations are allowed. For example, the fundamental
representation, which has dimension $N_c$, is constructed by
occupying an $N_c$-fold degenerate multiplet with a single ``slave"
fermion\cite{read}. For general $N_c$, with the exception of
$N_c=2$, this representation is not self-conjugate or, in other terms,
it is not particle-hole symmetric. Other representations can be
constructed\cite{withoff} by occupying the multiplet with $Q_f$ slave
fermions. For $Q_f=N_c/2$, which is available for $N_c$ even,
particle-hole symmetry is exact. We will see below that particle-hole
symmetry (self-conjugation) is a case of special interest. Notice that all
choices of representation are, in principle, valid extensions from the
physical $SU(2)$-invariant system.
Similar caveats have to  be made about the  choice of a particular
generator
in the algebra of $SU(N_c)$ that will represent the Zeeman term for
$N_c>2$.
In fact, in ref.~\cite{withoff} it was shown that some care has to be
taken in this choice in order to describe  a smooth weak-to-strong field
crossover.
In any event, we are only interested in
the extrapolation of the results at $N_c>2$ down to $N_c=2$ where there is
no ambiguities but they are present for all $N_c>2$.

In reference ~\cite{paperII} it was shown that, after integrating out
the fermion and impurity degrees of freedom, the impurity contribution
to the effective action $S_{eff}\equiv \beta F_{\rm imp}$ takes the form
\begin{eqnarray}
F_{\rm imp}& =& - {\frac{1}{\beta}}\sum_{\sigma=1}^{N_c} {\rm T}r \ln
\left[\partial_\tau + \epsilon_f + \sum_{l=1}^{N_f}
|\phi_{l}|^2 G_0(z) \right]
+ \int d\tau \left( \frac{N_c}{J_0}\left( \sum_{l=1}^{N_f} |\phi|^2\right)
- Q_f\epsilon_f\right)\nonumber\\
&\equiv& {\bar F}_{\rm imp}+
\int d\tau \left( \frac{N_c}{J_0}\left( \sum_{l=1}^{N_f} |\phi|^2\right)
- Q_f\epsilon_f\right)
\label{form3}
\end{eqnarray}
where $\phi_{l}$ are the Hubbard-Stratonovich fields introduced to
decouple the impurity in the large $N$ formalism. The properties of the
normal excitations is encoded in the function $G_0 (z)$ (where the
complex number $z=\epsilon+i\lambda$ is the analytic extension of the
energy). With the new definition of the DOS of Eq.~\ref{form1}, the
function $G_0 (z)$, defined by
\begin{equation}
G_0(z)~\equiv~-~\int_{-\infty}^{\infty}
\frac{d\epsilon}{2\pi v~v'}~\frac{\rho(\epsilon)}
{\epsilon - z}
\label{form2}
\end{equation}
now takes the form
\begin{equation}
{\rm Re} G_0(\epsilon + i\lambda) \equiv
\left\{
\begin{array}{l}
- \frac{\epsilon}{\pi v~v'}~\ln\left| \frac{\Delta_0}{\epsilon} \right|
\qquad {\rm for} \qquad |\epsilon| < \Delta_0 \cr
0 \qquad {\rm for} \qquad |\epsilon| > \Delta_0
\end{array}
\right.
\label{form4}
\end{equation}
\begin{equation}
{\rm Im} G_0(\epsilon + i\lambda) \equiv
\left\{
\begin{array}{l}
- \frac{|\epsilon|}{2 v~v'}~{\rm sgn}(\lambda)
\qquad {\rm for} \qquad |\epsilon| < \Delta_0 \cr
-~\frac{\Delta_0}{2 v~v'}~{\rm sgn}(\lambda) \qquad {\rm for}
\qquad |\epsilon|
>
\Delta_0
\end{array}
\right.
\label{form5}
\end{equation}
We define
\begin{equation}
\Delta~\equiv~\frac{\sum_{l=1}^{N_f}|\phi_l|^2}{\pi v~v'}
\label{form6}
\end{equation}
where $N_f$ is the number of ``flavors". For the problem of the d-wave
superconductor, $N_f$ in principle is the number of nodes and $N_f=4$.
However, we showed above that the {\it inter}-node couplings are always
different (and smaller) than the {\it intra}-node coupling. this
coupling anisotropy reduces to {\it one} the number of effective
flavors. Hence, from now on, we will set $N_f=1$.

At finite temperature $T$, the effective action of Eqn.(\ref{form3}) becomes an
(infinite) series running over (imaginary) Matsubara frequencies. Using this 
approach~\cite{doniach} the effective free energy becomes 
\begin{eqnarray}
{\bar F}_{\rm imp} & = & {\frac{N_c}{2\pi i}}
\int_{-\infty}^{\infty} \; d\epsilon \;
~{\frac{e^{\eta\epsilon}}{e^{\beta\epsilon}~+~1}}
~\ln\left[
{\frac{-(\epsilon+i\lambda) + \epsilon_f + (\sum_{\ell}^{}|\phi|^2)G_0(
\epsilon+i\lambda)}
{-(\epsilon-i\lambda) + \epsilon_f + (\sum_{\ell}^{}|\phi|^2)G_0(
\epsilon-i\lambda)}}
\right]
\nonumber\\
& \equiv & {\frac{N_c}{\pi}} \int_{-\infty}^{\infty} d\epsilon \;
n(\epsilon)
\;
\delta(\epsilon)
\label{eq:finite1}
\end{eqnarray}
where $\delta(\epsilon)$ is the {\it phase shift}~\cite{doniach},
$n(\epsilon)$ is the Fermi function
\begin{equation}
n(\epsilon)~=~
\frac{1}{e^{\beta \epsilon}+1}
\label{fermi}
\end{equation}
Explicitly we find
\begin{equation}
\delta(\epsilon) \equiv
\tan^{-1}
\left(
\frac{\lambda~+~\frac{\pi|\epsilon|\Delta}{2}}
{\epsilon~+~\epsilon\Delta\ln \left| \frac{\Delta_0}{\epsilon}\right|
~-~\epsilon_f}
\right)
\qquad\qquad
\left( \lambda \to 0^{+} \right)
\label{eq:finite4}
\end{equation}
$G_0(z)$ has a branch cut and the jump of
the function across this cut is energy dependent. This is an important
difference
with the usual Kondo effect in which the jump across the cut for
the function
$G_0(z)$ (see for example reference \cite{read}) is energy independent
and gives essentially the (constant) width of the
resonance. This will not be the case any longer as the
width of the resonance now becomes energy dependent.
This marks an important departure from the ``local Fermi liquid" (or the
resonant level model)
behavior~\cite{nozieres,newns} characteristic of the usual marginal Kondo
systems.

The large-$N_c$ analysis of this problem proceeds in the usual manner.
Given the impurity free energy ${\rm F}_{\rm imp}$, a set of values of
$\epsilon_f$ and $\Delta$ that minimize this free energy are sought. The
extremal values of $\epsilon_f$ and $\Delta$ satisfy the saddle point
equations (S.~P.~E.~)
\begin{equation}
\frac{\partial F_{\rm imp}}{\partial \Delta}~=~0
\qquad {\rm and} \qquad
\frac{\partial F_{\rm imp}}{\partial \epsilon_f}~=~0
\end{equation}
In the next subsection we will write explicit expressions for the SPE's
and solve them.
Thermodynamic magnitudes such as
the impurity entropy $S_{\rm imp}$, the impurity contribution to
the specific heat $C_{\rm imp}$ and the impurity contribution to the
susceptibility $\chi_{\rm imp}$ as functions of temperature (and magnetic
field)
can be computed from
the thermodynamic formulas
\begin{equation}
S_{\rm imp}~=~-~\frac{\partial F_{\rm imp}}
{\partial T},
\kern1cm
{\bar C}_{\rm imp}
~=~-~T~\frac{\partial^2 {\bar F}_{\rm imp}}{\partial T^2},
\kern1cm
\chi_{\rm imp}~=~-~\frac{\partial^2 {\bar F}_{\rm imp}}{\partial H^2}
\label{eq:finite7}
\end{equation}
where, the total impurity contribution to the free energy $F_{\rm imp}$
is given by
\begin{equation}
F_{\rm imp}~=~{\bar F}_{\rm imp} \left(\Delta, \epsilon_f, H, T
\right)
~+~\pi~\frac{v_F^2 N_c}{J_0}~\Delta
~-~Q_f\epsilon_f
\label{eq:finite8}
\end{equation}
Since both $\Delta$ and $\epsilon_f$ are also functions of $T$
and $H$, care must be taken to account for their contribution.
However, since $\Delta$ and $\epsilon_f$ satisfy the SPE's, we get
\begin{equation}
\frac{\partial F_{\rm imp} }{\partial T}~=~
\frac{\partial {\bar F}_{\rm imp}}{\partial T}\Big|_{\Delta,\epsilon_f}
\qquad {\rm and} \qquad
\frac{\partial F_{\rm imp} }{\partial H}~=~
\frac{\partial {\bar F}_{\rm imp}}{\partial H}\Big|_{\Delta,\epsilon_f}
\label{eq:finite9}
\end{equation}
Thus, only the explicit dependence on $T$ and $H$ matters.

\section{Saddle Point Equations}
\label{sec:spe}

Using the formalism of the previous section, the Saddle Point Equations
(SPE)
take the form
\begin{equation}
Q_f={\frac{1}{\pi}}
\int_{-D}^{+D}\;  d\epsilon \; n(\epsilon) \;
{\frac{\partial \delta}{\partial \epsilon_f}}(\epsilon)
\label{eq:spe1}
\end{equation}
and
\begin{equation}
N_c{\frac{\pi v_F^2}{J_0}}=
-{\frac{1}{\pi}}
\int_{-D}^{+D} \; d\epsilon \;
n(\epsilon)\;
{\frac{\partial \delta}{\partial \Delta}}(\epsilon)
\label{eq:spe2}
\end{equation}
where $D$ is the bandwidth cutoff.
In general we will be interested in the regime $T, H << \Delta_0 < D$.
In this regime, the contributions to the SPE from energies higher than
$\Delta_0$ can be well approximated by setting $T=H=0$. This amounts to
setting the Fermi function to be $n(\epsilon) \approx 0$, for $\Delta_0
\leq \epsilon \leq D$, and $n(\epsilon) \approx N_c$, for
$-D \leq \epsilon \leq -\Delta_0$. The SPE's thus are a sum of two
terms, one coming from energies $|\epsilon| \leq \Delta_0$ and one from
$D\geq |\epsilon| \geq \Delta_0$.

\subsection{Impurity Occupation}
\label{subsec:spe1}

The Saddle Point Equation Eq.~\ref{eq:spe1} reduces to
\begin{equation}
Q_f~=~\int_{-\Delta_0}^{\Delta_0} \frac{d\epsilon}{\pi}~n(\epsilon)~
\frac{ \frac{\pi}{2}|\epsilon|\Delta}
{\left(\frac{\pi}{2}|\epsilon|\Delta\right)^2 + \left(
\epsilon+\epsilon\Delta\ln \left|\frac{\Delta_0}{\epsilon}\right|
-\epsilon_f\right)^2}
+
\frac{N_c}{2}\left[ 1 - \frac{2}{\pi} \arctan
\left( \frac{\Delta_0 +\epsilon_f}
{\frac{\pi}{2}\Delta_0\Delta}\right)\right]
\label{eq:spe3}
\end{equation}

For the reminder of this paper, we will be interested in the physics of
this system close to the critical coupling constant. In that regime, the
singlet amplitude $\Delta$ becomes very small and the asymptotic
behavior of the SPE's in this domain can be evaluated explicitly.

Thus, close enough to the phase transition, where $\Delta$ is very small,
the
contribution from the last term in \ref{eq:spe3} becomes
\begin{equation}
\lim_{\Delta\to 0}
\left[ 1 - \frac{2}{\pi} \arctan
\left( \frac{\Delta_0 +\epsilon_f}
{\frac{\pi}{2}\Delta_0\Delta}\right)\right]
~\sim~ \left( \frac{\Delta_0}{\Delta_0+\epsilon_f}\right)\Delta
- \frac{2}{3\pi}\left( \frac{\frac{\pi}{2} \Delta_0\Delta}
{\Delta_0+\epsilon_f}
\right)^3 + \ldots
\label{eq:sp4}
\end{equation}
provided $\epsilon_f +\Delta_0 > 0$.

For the remainder of this section we will consider the Saddle Point
Equation in the case $T=H=0$. In this case, the SPE takes the form
\begin{equation}
I = \frac{N_c}{\pi} \int_{e^{1/\Delta}}^{\infty} \frac{dz}{z}~
\frac{\frac{\pi\Delta}{2}}
{\left(\frac{\pi\Delta}{2}\right)^2 + \left( \Delta\ln z +
\nu e^{-1/\Delta} z
\right)^2}
\label{eq:spe5}
\end{equation}
where $z \equiv e^{1/\Delta}~\frac{\Delta_0}{\epsilon}$
and $\nu \equiv \frac{\epsilon_f}{\Delta_0}$.
It is clear that the integrand in Eqn.(\ref{eq:spe5}) shows a crossover
behavior at
\[ \Delta \ln z_0 = \nu~e^{-1/\Delta}~z_0. \]
In the regime $\nu << \Delta << 1$, this equation has a root at large $z$,
given by
\[ z_0 \approx \left( \frac{\Delta}{\nu}~e^{1/\Delta}\right) \ln
\left(\frac{\Delta}{\nu}~e^{1/\Delta}\right) > e^{1/\Delta}
\]
Taking into account the change in the behavior of its denominator,
Eqn.(\ref{eq:spe5}) can be re-written in two pieces with the asymptotic
form
\begin{equation}
I^{<} \approx \frac{N_c}{2}~\left[ 1-
\frac{1}{1+\Delta\ln\frac{\Delta}{\nu}}
+ \frac{\Delta\ln\frac{1}{\Delta}}{\left(1+\Delta \ln
\frac{\Delta}{\nu}\right)^2}
\right] + \ldots
\label{eq:spe6}
\end{equation}
and
\begin{equation}
I^{>} \approx \frac{N_c}{\pi^2 \Delta} \ln\left[
1 + \frac{\left(\frac{\pi\Delta}{2}\right)^2}
{\left(1+\Delta \ln \frac{\Delta}{\nu}\right)^2}
\right] + \ldots
\label{eq:spe7}
\end{equation}
Getting everything together, Eqn.(\ref{eq:spe1}) reduces to the
expression
\begin{eqnarray}
Q_f &\approx & \frac{N_c}{2}\frac{\Delta}{1+\nu} + \frac{N_c}{2}
\frac{\Delta\ln\frac{\Delta}{\nu}}
{1+\Delta\ln\frac{\Delta}{\nu}}
+ \frac{N_c}{2} \frac{\Delta\ln\frac{1}{\Delta}}
{\left(1+\Delta\ln\frac{\Delta}{\nu}\right)^2}
\nonumber \\
&+ & \frac{N_c}{\pi^2\Delta}~\ln\left[ 1+
\frac{\left(\pi\frac{\Delta}{2}\right)^2}
{\left(1+\Delta\ln\frac{\Delta}{\nu}\right)^2}\right]
+ \ldots
\label{eq:spe8}
\end{eqnarray}
It is important to stress that, regardless of the approximations made in
evaluating the integrals, the SPE1 of Eq.~\ref{eq:spe1} is a relation
between $\Delta$ (the amplitude of the singlet) and the impurity Fermi
level (in units of the gap $\Delta_0$) $\nu$ at fixed occupation $Q_f$.
This relation is independent of the coupling constant and it must be
solved first. For the problem that we are discussing here, the relation
between $\nu$ and $\Delta$ is {\it singular}, as implied by the
logarithmic singularities in Eq.~\ref{eq:spe8}. We will see below that,
due to the presence of this singularity, the impurity Fermi level
$\epsilon_f$
is no longer simply related to the singlet amplitude $\Delta$.
This phenomenon does not occur in the conventional Kondo problem in
metals,
where the DOS is constant. It occurs for systems with a DOS vanishing
{\it faster} than linear with the energy. In this sense, the case of a
linear
DOS is a {\it marginal} system.

Let $x$ be the impurity filling fraction, $x=Q_f/N_c$.
The solution of Eq.~\ref{eq:spe8} takes the form
\begin{equation}
\nu(x,\Delta)={\sqrt{e}} \; \exp \left(-{\frac{1}{\Delta}} \;
{\frac{2x}{1-2x+\Delta}}+ {\frac{1}{1-2x+\Delta}}\right)
\left[1+O(\Delta,\Delta \ln \Delta)\right]
\label{eq:nu}
\end{equation}
where $e=2.7172\ldots$.

Hence, for generic values of $x=Q_f/N_c$,  the impurity Fermi level $\nu$
depends on the singlet amplitude $\Delta$ through an essential singularity
of the form $ \exp(-{\rm const}/\Delta)$. As $Q_f \to {\frac{N_c}{2}}$
($x \to {\frac{1}{2}}$), there is a crossover in the functional form
of $\nu$
which now behaves like $\exp(-{\rm const}/\Delta^2)$, which vanishes much
faster
as $\Delta$ approaches zero. It is interesting to note that if the
contributions
from the states  with energies between $-D$ to $-\Delta_0$ had been
neglected altogether, $\nu$ would have {\it vanished} identically at
$Q_f={\frac{N_c}{2}}$, for all finite values of $\Delta$.
Since {\it at} $Q_f={\frac{N_c}{2}}$ the hamiltonian has an
exact particle-hole symmetry, it may appear that
$\nu=\epsilon_f/\Delta_0$ should have to vanish exactly at this point.
In fact, it does not vanish due to states whose DOS violate the strict
linear behavior of the DOS at low energies.

\subsection{Equation of State}
\label{subsec:spe2}

Let us consider now the second SPE, Eq.(\ref{eq:spe2}). This equation
relates $\Delta$ (the amplitude of the singlet) to the coupling constant
(once the relation between $\epsilon_f$ and $\Delta$ is known). We will
regard this equation as an {\sl equation of state}.

At $T=0$ and $H=0$ the SPE2, Eq.(\ref{eq:spe2}) can be written as
\begin{eqnarray}
\frac{\pi^2 v_F^2}{J_0}
& = &
\int_0^{\Delta_0} d\epsilon \frac{\pi}{2}~
\frac{\epsilon(\epsilon +\epsilon_f)}
{\left(\frac{\pi\epsilon\Delta}{2}\right)^2
+
\left(\epsilon+\epsilon_f+\epsilon\Delta \ln\frac{\Delta_0}{\epsilon}
\right)^2} \nonumber \\
& + &
\int_{\Delta_0}^{D} d\epsilon \frac{\pi}{2}~\Delta_0~
\frac{\epsilon(\epsilon+\epsilon_f)}
{\left(\frac{\pi\Delta_0\Delta}{2}\right)^2
+
\left( \epsilon+\epsilon_f
\right)^2
}
\label{eq:state1}
\end{eqnarray}
In the second integral of the r.h.s. of Eq.(\ref{eq:state1})
it is useful to perform the change of variables
$ u = \epsilon_f/\epsilon$ while,
in order to treat the first integral or the r.h.s. of Eq.(\ref{eq:state1})
we use again $ z = e^{1/\Delta}(\Delta_0/\epsilon)$. As above,
$\nu = \epsilon_f/\Delta_0$ and $\nu_D = \epsilon_f/D$.
We can write
\begin{eqnarray}
\frac{1}{g_0} & = &
\frac{\pi}{2}~e^{1/\Delta} \int_{e^{1/\Delta}}^{\infty}
\frac{dz}{z^2}~
\frac{\left( 1 + \nu z e^{-1/\Delta}\right)}
{
\left(\Delta\frac{\pi}{2}\right)^2 + \left(\nu~z e^{-1/\Delta}
+ \Delta \ln z \right)^2
}\nonumber \\
&& \qquad \qquad + \frac{\pi}{2} \int_{\nu_D}^{\nu}
\frac{du}{u}~\frac{1+u}{\left(\Delta\frac{\pi}{2}\right)^2
+ \left( 1+ u \right)^2}
\label{eq:state2}
\end{eqnarray}
where we have defined the dimensionless coupling constant $g_0$ by
\[
\frac{1}{g_0}~=~ \frac{1}{\Delta_0}~\frac{\pi^2 v_F^2}{J_0}
\]

The second integral on the r.h.s. of Eq.(\ref{eq:state2})
can be shown to give the leading contributions
\begin{equation}
\frac{\pi}{2}
\int_{\nu_D}^{\nu}
\frac{du}{u}~\frac{1+u}{\left(\Delta\frac{\pi}{2}\right)^2
+ \left( 1+ u \right)^2}
\sim
\frac{\pi}{2} \ln~\frac{D}{\Delta_0}~\frac{1}{1+ a^2}
- \frac{\pi}{2}~\nu \left(1 - \frac{\Delta_0}{D}\right) (1- 7 a^2)
+ \ldots
\label{eq:state3}
\end{equation}
where $a^2 \equiv (\pi\Delta/2)^2 << 1$ in accordance to the
hypothesis that $\Delta$ is small in the regimes in which we are
interested.
The first integral of the r.h.s. of Eq.(\ref{eq:state2}) is
treated in the Appendix as an example of the approximations used.
Retrieving here the results of Eq.(\ref{eq:app8}) we write
Eq.(\ref{eq:state2}) in the form
\begin{equation}
\frac{1}{g_0} ~=~ \frac{\pi}{2} \ln\left(\frac{D}{\Delta_0}\right)
~+~ \frac{\pi}{2}~\frac{1}{1+ \left(\pi\Delta/2\right)^2} ~-~
\pi~\frac{\Delta}{\left(1+ \left(\pi\Delta/2\right)^2\right)^2}
~+~O\left(\Delta^2\right)~+~\ldots
\label{eq:state4}
\end{equation}
Now we define the {\it critical coupling constant} as the limit for
$\Delta \to 0$ of Eq.(\ref{eq:state4})
\begin{equation}
\frac{1}{g_c} ~=~\frac{\pi}{2} \ln\left(\frac{D}{\Delta_0}\right)
~+~ \frac{\pi}{2}
\label{eq:state5}
\end{equation}
For small $\Delta$ we obtain the scaling equation
\begin{equation}
\frac{1}{g_c}~-~\frac{1}{g_0}~=~ \pi~\Delta ~+~ \frac{\pi}{4}
\left(\frac{\pi\Delta}{2}\right)^2\left[ 3 - \left(\frac{\Delta_0}{D}\right)^2
\right] ~+~ \ldots
\label{eq:state6}
\end{equation}

\section{Impurity magnetic susceptibility}
\label{sec:chi}

In order to consider the effect of a magnetic field in our model it
is necessary to proceed with some care.
In principle we need to go back to the original model to look into
the effects of a finite $H$.
In terms of the Nambu-Gorkov spinors the ``free hamiltonian" $H_0$ now
becomes
\begin{equation}
H_0=\sum_{{\vec k}}^{} \Phi^{\dagger}({\vec k})
\left[ \left( \epsilon({\vec k}) + \mu\right)\tau_3
- \Delta({\vec k}) \tau_1 - H\right] \Phi({\vec k})
\label{eq:magf1}
\end{equation}
Here the magnetic field $H$ is multiplied by the
$2\times 2$ identity matrix.
The consequence of the introduction of a finite magnetic field is thus,
the generation of a finite relative shift in
the zero point for the energy, but not necessarily a finite
density of quasi-particle states
within the nodes of the gap (however, see below).
The eigenfunctions remain unchanged but the eigenenergies are shifted by
$H$
\begin{equation}
{\rm E}~=~ -~H~\pm \sqrt{ \epsilon^2(\vec k ) ~+~ \Delta^2 (\vec k) }
\label{eq:magf2}
\end{equation}
Thus, after the expansion in small momentum around the {\sl nodes}
of the gap, the two-dimensional spinors will disperse with
\begin{equation}
{\rm E} ~=~ -~H~\pm~\sqrt{v v~'} p
\label{eq:magf3}
\end{equation}
where $p$ has been defined before, in the model without field.

It is not difficult to convince oneself, by going through the (several)
transformations involved in the reduction to the effective one-dimensional
model that, at the level of the one-dimensional hamiltonian, the
magnetic field enters as a {\sl true} magnetic field coupled now, to the
one-dimensional chiral fermions. Thus,
\begin{eqnarray}
H_{eff} & = &
\sum_{\ell=1}^{4} \sum_{\sigma =\uparrow, \downarrow}
\int_{-\infty}^{\infty}
\frac{dp}{2\pi} \left(
{\sqrt{vv'}} p ~-~H\tau_3\right)
d^{\dagger}_{\ell\sigma}(p) d_{\ell\sigma}(p)
\nonumber \\
& + &
\sum_{\ell=1}^{4} \sum_{\sigma,\nu =\uparrow, \downarrow}
J_{\ell}~ \left[ \int_{-\infty}^{\infty} \frac{dp}{2\pi} \sqrt{|p|}
d^{\dagger}_{\ell\sigma}(p)
\right] {\vec \tau}_{\sigma\nu} \cdot {\vec S}_{\rm imp}
\left[ \int_{-\infty}^{\infty} \frac{dp'}{2\pi} \sqrt{|p'|}
d_{\ell\nu}(p')
\right]
\label{eq:magf4}
\end{eqnarray}
The change in the kinetic energy, depending on the spin polarization,
changes the form of the function $G_0(\omega,H)$
\begin{eqnarray}
G_0(\omega, H) & = & \int_{-\infty}^{\infty} \frac{dp}{2\pi}~
\frac{|p|}{\omega - \sqrt{vv'}p + H\tau^3}
\nonumber \\
& = &
\frac{1}{2\pi vv'} \int_{-\infty}^{\infty}
\frac{|\epsilon|~d\epsilon}{(\omega+H)-\epsilon}~\frac{1}{2}\left(1+\tau_3
\right) ~+~
\frac{1}{2\pi vv'} \int_{-\infty}^{\infty}
\frac{|\epsilon|~d\epsilon}{(\omega-H)-\epsilon}~\frac{1}{2}\left(1-\tau_3
\right)
\end{eqnarray}
where $\tau_3$ represents an $SU(N_c)$ diagonal generator having
$r$ elements with eigenvalue $+1$ and $N_c-r$ elements with
eigenvalue $-1$. In what follows we will take $r=N_c/2$ which respects
the $H \to -H$ symmetry of the $SU(2)$ theory. For general $r$, a
particle-hole transformation {\it is not} equivalent to $H \to -H$. Bout
for $r=N_c/2$ these symmetry transformations are equivalent. In other
terms, for general $r$ this magnetic field breaks both the $H \to -H$
symmetry {\it and} particle-hole.
Notice, however, that the
{\it representation} of the impurity is determined solely by the charge
$Q_f$ and it is unrelated to $r$.
In the presence of the field, the impurity level has an effective
filling factor $2Q_f/N_c$. We will see below that $Q_f=N_c/2$
is a special case. For the physical case $N_c=2$ there is only one
possible representation ({\it i.~e.\/} spin $S=1/2$) which corresponds
to $Q_f=1=N_c/2$. For general $N_c$ these two situations do not
necessarily
coincide.

These changes will be reflected in the phase shift defined in section~
\ref{sec:largeN}. In presence of a finite field the impurity free energy
${\bar F}_{\rm imp}$ can be written as
\begin{equation}
{\bar F}_{\rm imp}  \equiv
\frac{N_c}{2\pi}~
\int_{-\infty}^{\infty}
d\epsilon~n(\epsilon)~\left[ \delta
(\epsilon+H) ~+~ \delta(\epsilon-H)\right]
\label{eq:newg1}
\end{equation}
where $\delta(\epsilon)$ is given by Eq.(\ref{eq:finite4})
and $n(\epsilon)$
is the Fermi function.
Eq.(\ref{eq:newg1}) is manifestly invariant under
the transformation $H \to -~H$.

The magnetization and the susceptibility are given respectively by
\begin{equation}
M_{\rm imp}~=~ -~\frac{\partial {\bar F}_{\rm imp}}{\partial H};
\kern1in
{\chi}_{\rm imp}~=~-~\frac{\partial^2 {\bar F}_{\rm imp}}{\partial H^2}
\label{eq:chi2}
\end{equation}
which take the form
\begin{equation}
M_{\rm imp}  =
{\frac{N_c}{2\pi}} \int_{-\infty}^{+\infty} d\epsilon \left(
~{\frac{\partial n}{\partial \epsilon}} (\epsilon-H)~-~
{\frac{\partial n}{\partial \epsilon}}(\epsilon+H)\right)\delta(\epsilon)
\label{eq:chi3}
\end{equation}
In the limit $T\to 0$, the function $\frac{\partial n(\epsilon)}
{\partial \epsilon}$ approaches a negative Dirac $\delta$-function
localized at
$\epsilon = 0$. In this limit we find
\begin{equation}
 M_{\rm imp} (0,H) ~=~-\frac{N_c}{2\pi}
\left[ \delta(H)-\delta(-H)\right]
\label{eq:chi4}
\end{equation}
Now we can use Eq.(\ref{eq:finite4}) to write an explicit expression
for the magnetization
\begin{equation}
 M(0,H) =
-\frac{N_c}{2\pi} \tan^{-1}\left(
\frac{ \frac{\pi}{2} H \Delta }{H+H\Delta
\ln\frac{\Delta_0}{H}-\epsilon_f}
\right)
- \frac{N_c}{2\pi}
\tan^{-1}\left(
\frac{ \frac{\pi}{2} H \Delta }{H+H\Delta \ln\frac{\Delta_0}{H}+
\epsilon_f} \right)
\label{eq:chi5}
\end{equation}
It is easy to see that in the limit $H<<\epsilon_f(0)$, one has
\begin{equation}
M_{\rm imp}~\sim~\frac{N_c}{2}\frac{\Delta(0)}{\epsilon_f^2(0)}
~H^2\left(1+\Delta(0)\ln\frac{\Delta_0}{H}\right)
\label{eq:chi6}
\end{equation}
This expression shows that the impurity contribution to the magnetization
vanishes as $H^2\ln H$ with $H\to 0$. As expected, the impurity
magnetization vanishes as the field goes to zero thus showing that the
magnetic impurity has been screenined. However, in a conventional {\it
marginal} Kondo system, the magnetization vanishes {\it lineraly} with
the field. Here instead we find a faster field dependence .

It can be shown, using similar arguments, that a general expression for
the impurity contribution to the magnetic susceptibility
is given by
\begin{eqnarray}
\chi_{\rm imp}(T,H)&=&{\frac{N_c}{2\pi}}\int_{-\infty}^{+\infty}
{\frac{\partial n}{\partial \epsilon}}
\left[{\frac{\partial \delta}{\partial \epsilon}}(\epsilon+H)+
{\frac{\partial \delta}{\partial \epsilon}}(\epsilon-H)
\right]
\nonumber\\
&=& -{\frac{N_c}{2\pi}}
\int_{-\infty}^{+\infty} dx {\frac{e^x}{(e^x+1)^2}}
\left[{\frac{\partial \delta}{\partial \epsilon}}\Big|_{xT+H}+
{\frac{\partial \delta}{\partial \epsilon}}\Big|_{xT-H}
\right]
\label{eq:chi7b}
\end{eqnarray}
At zero temperature the susceptibility becomes
\begin{equation}
\chi_{\rm imp}(0,H)~=~-~\frac{N_c}{2\pi}
\left[ \frac{\partial\delta}{\partial\epsilon}\Big|_{-H}
+ \frac{\partial\delta}{\partial\epsilon}\Big|_{H}\right]
\label{eq:chi7}
\end{equation}
Thus, we find that the susceptibility at zero
temperature and at low fields ( $H\ll \epsilon_f(0)$) is
\begin{equation}
\chi_{\rm imp} (0,H)  =  {\frac{
N_c\Delta
\left\{
\epsilon_f^2 \left(H+H\Delta\ln\frac{\Delta_0}{H}\right)
-
\frac{1}{2} H \Delta
\left[
\left(\frac{\pi H \Delta }{2}\right)^2+
\left(H+H\Delta\ln\frac{\Delta_0}{H}\right)^2
+\epsilon_f^2
\right]
\right\}}
{\left( \left(\frac{\pi H \Delta }{2}\right)^2+
\left(H+H\Delta\ln\frac{\Delta_0}{H}\right)^2
+\epsilon_f^2\right)^2 -
4\epsilon_f^2\left(H+H\Delta\ln \frac{\Delta_0}{H}\right)^2
}}
\label{eq:chi8}
\end{equation}
It should be noticed that in all of these expressions, the quantities
$\epsilon_f$
and $\Delta$ are functions of the field $H$, with a limiting value
$\epsilon_f(0)$ and
$\Delta(0)$ for $T=H=0$ found in section \ref{sec:spe}.
The quantity $\Delta(0)$ should not be confused with $\Delta_0$.
Hereafter we will set $\Delta=\Delta(0)$ and $\epsilon_f=\epsilon_f(0)$.
The magnetic susceptibility obtained from Eq.(\ref{eq:chi6}) agrees
with the
limit $\epsilon_f>>H$ of Eq.(\ref{eq:chi8}) and gives
\begin{equation}
\chi_{\rm imp}~\sim~
N_c \left(\frac{\Delta}{\epsilon_f}\right)^2~H\ln
\frac{\Delta_0}{H}
~+~ N_c~\frac{\Delta}{\epsilon_f^2}~H
\left(1-\frac{\Delta}{2}\right)~+~\ldots
\label{eq:chi9}
\end{equation}
In the opposite regime, $H\ll T \ll \epsilon_f$, the susceptibility is
\begin{equation}
\chi_{\rm imp} (T,0) \approx 2 N_c \ln 2  \;
\left({\frac{\Delta}{\epsilon_f}}\right)^2 T \ln ({\frac{\Delta_0}{T}})
\label{eq:chilow}
\end{equation}
To summarize, we find that in the low field
limit the zero temperature magnetization vanishes like
$H^2 \ln (\Delta_0/H)$. However,
in contrast with the conventional ``Fermi liquid" behavior of the Kondo
effect in metals, the magnetic susceptibility is {\it also} found to
vanish in the low field limit as $H \ln(\Delta_/H)$ and at zero
temperature. Hence, in this
non-marginal Kondo system, the magnetic impurity is {\it overscreened}
even for a single channel of fermions. In the low temperature, zero
field regime, the impurity susceptibility has a $T \ln (\Delta_0/T)$
behavior which again shows that the impurity is overscreened.

\section{Impurity entropy ans specific heat}
\label{sec:ent}

We can estimate the impurity contribution to the specific heat
in the limit $T<<H$, in the screening regime.
Using the SPE and some straightforward algebra,
the impurity contribution to the entropy is
\begin{equation}
S_{\rm imp}~=~-~\frac{\partial F}{\partial T}\Big|_{\epsilon_f,\Delta}
~=~\frac{N_c}{2\pi}\int_{-\infty}^{\infty} d\epsilon~\frac{\epsilon}{T}
{\frac{\partial n}{\partial \epsilon}}
\left[ \delta(\epsilon+H) + \delta(\epsilon-H)\right]
\label{eq:ent1}
\end{equation}
Using the scaling $\epsilon=xT$ this is
\begin{equation}
S_{\rm imp}~=~-~\frac{N_c}{2\pi}
\int_{-\infty}^{\infty} dx~x~\frac{e^x}{(e^x+1)^2}
\left[ \delta(xT+H)+\delta(xT-H)\right]
\label{eq:ent2}
\end{equation}
In the limit $T << H$, we may expand the phase shift around
the point $x=0$ to get
\begin{equation}
S_{\rm imp}~=~\chi_{\rm imp}(0,H)~T~\int_{-\infty}^{\infty}
dx~x^2~\frac{e^x}{(e^x+1)^2}
\label{eq:ent3}
\end{equation}
where $\chi_{\rm imp}(0,H)$ has been obtained in the previous
section.
The impurity contribution to the specific heat is then, equal to the
impurity contribution to the entropy in this limit.
Our result shows that the impurity entropy vanishes at $T\to 0$
and finite field, for $T<<H$.

The general form of the specific heat is
\begin{eqnarray}
C_{\rm imp}
& = & {\frac{N_c}{2\pi T}}~\int_{-\infty}^{+\infty} d\epsilon~\epsilon^2~
{\frac{\partial n}{\partial \epsilon}}~\left[{\frac{\partial\delta}
{\partial\epsilon}} (\epsilon+H) +
{\frac{\partial\delta}{\partial\epsilon}} (\epsilon-H)\right]
\nonumber \\
& = &-{\frac{N_c T}{2\pi}}~\int_{-\infty}^{+\infty} dx\; x^2
{\frac{e^x}{(e^x+1)^2}}
\left[{\frac{\partial\delta}{\partial\epsilon}}\Big|_{xT+H}+
{\frac{\partial\delta}{\partial\epsilon}}\Big|_{xT-H}
\right]
\label{eq:ent4}
\end{eqnarray}
In the regime $H\ll T \ll \epsilon_f$ we find
\begin{equation}
C_{\rm imp}(0,T) \approx 9 \zeta(3) N_c
{\frac{\Delta^2}{\epsilon_f^2}} T^2 \ln({\frac{\Delta_0}{T}})
\label{eq:Clow}
\end{equation}
where $\zeta(3)$ is the Riemann $\zeta$-function at $3$ and it is a
number of the order of unity.

Using the results of Eq.~(\ref{eq:chilow}) and
Eq.~(\ref{eq:Clow}), we can compute the Wilson Ratio for the
regime $H\ll T \ll \epsilon_f$ and find
\begin{equation}
{\frac{C_{\rm imp}(H,T)}{T \chi_{\rm imp}(H,T)}}
\approx {\frac{9 \zeta(3)}{2 \ln 2}}
\label{eq:Wlow}
\end{equation}
It is interesting that the ratio is still finite in spite of the fact
the both the specific heat and the susceptibility behave very differently
than in a Fermi liquid.

In the high field limit $T \ll H \ll \epsilon_f$ the impurity specific
heat is
\begin{equation}
C_{\rm imp}(H,T) \approx  N_c {\frac{\pi^2}{3}}   \;\;
\left({\frac{\Delta}{\epsilon_f}}\right)^2 T~ H \ln ({\frac{\Delta_0}{H}})
\label{eq:Chigh1}
\end{equation}
which obeys the relation
\begin{equation}
C_{\rm imp}(H,T) \approx {\frac{\pi^2}{3}} T \chi_{\rm imp}(0,H)
\label{eq:Chigh}
\end{equation}
This result leads to a new Wilson Ratio
\begin{equation}
W={\frac{C_{\rm imp}(H,T)}{T \chi_{\rm imp}(H,0)}}={\frac{\pi^2}{3}}
\label{eq:Whigh}
\end{equation}
which is essentially identical to the Wilson Ratio for the Kondo effect in
Fermi liquids.

Hence we found that in the strong coupling Kondo phase, the impurity
specific heat at low temperature and low fields behaves like $T^2 \ln
(\Delta_0/T)$ and $T ~H \ln(\Delta_0/H)$ depending on whether $H\ll T$
or $T\ll H$. Only for $T\ll H$ we find the conventional linear $T$
behavior of the (zero field) specific heat of the (marginal) Kondo
effect in Fermi liquids. Notice however that the slope $\gamma$ of the
specific heat in this regime is field dependent and behaves like
$H\ln(\Delta_0/H)$. However, in spite of these differences,
we found that the conventionally defined
Wilson ratio is still finite but it is different in both regimes.

\section{Discussion}
\label{sec:discussion}

In this paper we constructed a model for the problem of a magnetic
impurity in a $d_{x^2-y^2}$ superconductor. We solved this problem using
the large-$N_c$ approximation and found that there is a quantum phase
transition from a phase in which the impurity is nearly free to a
phase in which it is overscreened. We estimated the value of the
critical coupling constant $J_c$. We found that $J_c$ could be both
smaller or larger than $\Delta_0$, the gap of the d-wave superconductor,
but it is certainly smaller than the bandwidth $D$ of the electrons that
participate in the superconductivity.

This result agrees with recent
work by K.~Ingersent\cite{ingersent} on a related system. Ingersent used
a wilsonian numerical RG approach and found that the critical coupling
runs off to the cutoff (strong coupling) unless either particle-hole
symmetry is broken (for the band fermions) or additional high energy
states with a flat DOS were added. In the problem of the d-wave
superconductor the former possiblity is excluded by the superconductivity
itself but the latter is required since such states are always there.
In any event there is no reason to require that $J_c$ should be smaller
than $\Delta_0$. In fact, even if $J_c \geq \Delta_0$, the Kondo scale
$T_K$ does not track $J_c$ and it is almost always smaller than
$\Delta_0$ (in fact, quite a bit smaller!).
The value of the
critical coupling constant is non-universal and it depends on details of
the high energy physics of the system. Thus, our approximations have
emphasized the role of the nodes and replaced the states above $\Delta_0$
by a ``flat band". Clearly, the solution of the saddle point equations
with  the full band structure of the hamiltonian of Eq.~\ref{eq:H} will
yield a different (possibly smaller) value of $J_c$. The same caveats
apply to the numerical RG calculation of K.~Ingeresent, in which a
specific discretization of the effective model is used. In fact, in most
of his work, Ingeresent uses Wilson's logarithmic discretization which
is very accurate for the Kondo problem in metals since it is tailored to
reproduce the logarithmic singularities at high energies of the
conventional (marginal) Kondo problem. In the case that we examine here,
the system is very far away for its ``lower critical dimension" .
This approach should {\it overestimate} $J_c$, probably by quite a bit.
In any event, the actual value of $J$ itself depends on microscopic
physics of the cuprates and there is no reason to believe that it should
be tied to $\Delta_0$.

We investigated in detail the thermodynamic behavior
(impurity susceptibility and  specific heat) in the overscreened phase
where we found that, the impurity susceptibility vanishes like
$H \ln (\Delta_0/H)$ (for $ T \ll H \ll T_K$) or $T \ln
(\Delta_0/T)$ (for $H \ll  T \ll T_K$) with a
crossover at $T \approx H$. For a Fermi liquid, the impurity susceptibiliy
approaches aconstant value ast $T \to 0$.
The specific heat , on the other hand, was found to vanish like
$T H \ln (\Delta_0/H)$ (for $ T \ll H \ll T_K$) or $T^2 \ln
(\Delta_0/T)$ (for $H \ll  T \ll T_K$) . In a Fermi liquid it vanishes
linearly with $T$. The change in the power law behavior is an extension
of the earlier work by Withoff and Fradkin\cite{withoff}. The additional
logarithmic singularity is an indication that $r=1$ is like an upper
critical dimension for the Kondo problem\cite{paperII}.
The interesting quantum critical behavior,
accessible for $T,H \gg T_K$ was not discussed here and will be the
subject of a separate publication\cite{paper4}.

There are several important effects that we have not included here. One is
the effect of {\it random} potential scattering which, na{\" \i}vely
may induce a non-zero DOS at the Fermi energy $E_F=0$. Even if this
effect is there, the effective DOS $N(E)$ is very small. In
\cite{fisher,dirac} it was shown that $N(E)\approx \exp(-{\rm const.}/w)$
(where $w$ is the witdth of the distribution) and that the elastic mean
free path is exponentially long, $\ell \sim \exp(+{\rm const.}/w)$.
Since at $J_c$ we have a transition
from a state with a {\it divergent} zero temperature susceptibilty
(Curie-like) to an overscreened state with {\it vanishing}
susceptibility, the rounding effects of a finite (but very small) DOS
should be a very small correction if the material is clean.
A more interesting, and perhaps more important, effect that was not
included here is the presence of explicit pair-breaking by the vanishing
of the amplitude of the d-wave order parameter at the impurity site.
This effect should give rise to interesting Andreev like processes which
may well alter the physics of this problem. We will discuss this
problem elsewhere\cite{paper4}. Finally, corrections to the $N_c \to
\infty$ limit remain to be estimated.

\section{Acknowledgements}
\label{sec:ack}

We have benefitted from very iluminating discussions on this subject with
Prof.~Kevin Ingersent who was also kind to share  his as-yet unpublished
results with us. 
We are also grateful to Prof.~Qimiao Si for many useful duscussions
and suggestions. 
EF wishes to thank Prof.~David Pines for many comments and
discussions on the role of magnetic impurities in d-wave superconductors.
CC wishes to thank Prof. John Berlinsky for informative discussions about
impurities in superconductors.
EF was a visitor at the Institute for Theoretical Physics of the
NSF at the University of California Santa Barbara during part of this work
in connection with its Program on Non-Fermi Liquids in Solids, organized
by P.~Coleman, B.~Maple and A.~Millis.
CC wishes to thank the organizers of the Conference on Non-Fermi Liquid
Physics for allowing him to participate and present a preliminary version
of this work.
We are grateful to the ITP and to its Director, Prof.~Jim Hartle, and to the
organizers of the Program for their kind hospitality.
This work was supported in part by the National Science Foundation through
the grants NSF-DMR-94-24511  at the Department of Physics of the University of
Illinois at Urbana-Champaign and NSF-DMR-89-20538/24 at the Materials Research
Laboratory of the University of Illinois at Urbana-Champaign. We are
particularly grateful to Prof.~Jim Wolfe for the MRL support of this work.


\newpage
\appendix

\section{Estimation of integrals}
\label{sec:B}
The integral of Eqn.(\ref{eq:state1}) is a good representative for the
approximations made in treating the integrals of section~\ref{sec:spe}.
The expression given in Eq.(\ref{eq:state3}) is easy to obtain.
As for the first integral of the r.h.s. of Eq.(\ref{eq:state2})
we have
\begin{equation}
\frac{\pi}{2}~e^{1/\Delta} \int_{e^{1/\Delta}}^{\infty}
\frac{dz}{z^2}~
\frac{\left( 1 + \nu z e^{-1/\Delta}\right)}
{
\left(\Delta\frac{\pi}{2}\right)^2 + \left(\nu~z e^{-1/\Delta}
+ \Delta \ln z \right)^2
}~\equiv~{\rm K}~+~{\rm I}
\label{eq:app1}
\end{equation}
where the splitting of the integral corresponds to the $+$ sign in
the numerator.
At the value $z_0 = \left(\frac{\Delta}{\nu}~e^{1/\Delta}\right)\ln z_0$
there is a crossover in the behavior of the denominator of the
integrand. For $ z < z_0$ the leading term is $\Delta \ln z$;
for $z > z_0$ we can keep the term $\nu~z~e^{-1/\Delta}$.
Also $z_0$ can be approximated by
\[ z_0 \approx \left( \frac{\Delta}{\nu}~e^{1/\Delta}\right) \ln
\left(\frac{\Delta}{\nu}~e^{1/\Delta}\right) > e^{1/\Delta}
\]
Then we can split the integrals as
\begin{equation}
{\rm I}^{\ <} ~=~
\frac{\pi}{2}~\frac{\nu}{\Delta^2}
\int_{e^{1/\Delta}}^{z_0} \frac{dz}{z}~
\frac{1}{\ln^2 z + \left(\frac{\pi}{2}\right)^2}
\label{eq:app2}
\end{equation}
The change of variables $t=\frac{2}{\pi}~\ln z$
makes the integration straightforward to give
\begin{eqnarray}
{\rm I}^{\  <} & = &
\frac{\nu}{\Delta^2} \left[
\arctan\left( \frac{2}{\pi}\ln z_0\right)
- \arctan \left( \frac{2}{\pi\Delta}\right)\right]
~ \sim ~
\nu~\frac{\pi}{2}~
\frac{\ln\left( \frac{\Delta}{\nu} \ln \left(\frac{\Delta}{\nu}
e^{1/\Delta}\right) \right)}
{1+ \Delta \ln
\left( \frac{\Delta}{\nu}\ln\left(\frac{\Delta}{\nu} e^{1/\Delta}\right)
\right) }
\label{eq:app3}
\end{eqnarray}
On the other hand
\begin{equation}
{\rm I}^{\  >} = \nu~\frac{\pi}{2}~\frac{1}{\nu^2 e^{-2/\Delta}}
\int_{z_0}^{\infty} \frac{dz}{z}~\frac{1}{z^2+\left(e^{1/\Delta}~
\frac{\pi\Delta}{2\nu}\right)^2}
\label{eq:app4}
\end{equation}
which can be integrated by partial fractions to give
\begin{equation}
{\rm I}^{\  >} = \nu~\frac{\pi}{2}~
\frac{1}{\left(1+\Delta\ln\frac{\Delta}{\nu}
\right)^2}
\label{eq:app5}
\end{equation}
Similarly we have
\begin{equation}
{\rm K}^{\  <} = \frac{\pi}{2}~e^{1/\Delta}~\frac{1}{\Delta^2}
\int_{1/\Delta}^{\ln z_0} e^{-t}~\frac{dt}{\left(\pi/2\right)^2 + t^2}
\label{eq:app6}
\end{equation}
and
\begin{equation}
{\rm K}^{\  >} = \frac{\pi}{2}~\frac{e^{3/\Delta}}{\nu^2}
\int_{z_0}^{\infty} \frac{dz}{z^2}~\frac{1}{z^2+\left(e^{1/\Delta}~
\frac{\pi\Delta}{2\nu}\right)^2}
~<~\frac{e^{1/\Delta}}{\nu}~\frac{1}{z_0}~{\rm I}^{>}
~\sim~\frac{\pi}{2} \nu~\frac{1}{\left(1+\Delta\ln\frac{\Delta}{\nu}
\right)^3}
\label{eq:app7}
\end{equation}
Hence, again ${\rm K}^{\  >}$, as it was the case with ${\rm I}^{\ >}$,
can be
neglected since its contribution is at least of the ${\rm o}(\nu)\sim
{\rm o} (e^{-1/\Delta}) << \Delta$.
The leading contribution comes from Eqn.(\ref{eq:app6}) which can be
recasted in terms of an exponential integral function and produces a
constant contribution and a linear term in $\Delta$.
The leading contribution for small $\Delta$ give
\begin{equation}
{\rm K}^{\  <} ~\sim~ \frac{\pi}{2}
\left[ \frac{1}{1+ \left(\pi\Delta/2\right)^2} ~-~
\frac{2\Delta}{\left(1+ \left(\pi\Delta/2\right)^2\right)^2}
~+~O(\nu)
~+~ \ldots\right]
\label{eq:app8}
\end{equation}
With these results, the equation of state, keeping only the leading
order contributions is
\begin{equation}
\frac{1}{g_0} ~=~ \frac{\pi}{2} \ln\left(\frac{D}{\Delta_0}\right)
~+~ \frac{\pi}{2}~\frac{1}{1+ \left(\pi\Delta/2\right)^2} ~-~
\pi~\frac{\Delta}{\left(1+ \left(\pi\Delta/2\right)^2\right)^2}
~+~O\left(\Delta^2\right)~+~\ldots
\label{eq:app9}
\end{equation}


\newpage

\begin{references}

\bibitem{pines} P.~Monthoux and D.~Pines,
{\sl Phys.~Rev.~Lett.~\/} {\bf 69}, 961 (1992).

\bibitem{wenger} A.~A.~Nersesyan, A.~M.~Tsvelik and F.~Wenger,
{\sl Phys.~Rev.~Lett.~\/} {\bf 72}, 2628 (1994);
{\sl Nucl.~Phys.~\/} {\bf B438}, 561 (1995).

\bibitem{balatsky} A.~V.~Balatsky, M.~I.~Salkola and  A.~Rosengren,
{\sl  Phys.~ Rev.\/} {\bf B 51}, 15547 (1995).

\bibitem{franz} M.~Franz, C.~Kallin and A.J.~ Berlinsky,
{\sl  Phys.~ Rev.\/} {\bf B 54}, 6897 (1996).

\bibitem{pair} K.~Ueda and T.~M.~Rice in {\sl Theory of Heavy Fermions
and Valence Fluctuations}, T.~Kasuya and T.~Saso editors
(Springer-Verlag, Berlin 1985), p.~267; L.~P.~Gorkov and P.~ A.~Kalugin,
{\sl Pisma ZhETP}{\bf 41}, 208 (1985)[ {\sl JETP.~Lett.\/}{\bf 41},
253 (1985).

\bibitem{nmr-pines}A.~V.~Balatsky, P.~Monthoux and D.~Pines,
{\sl  Phys.~ Rev.\/} {\bf B 50}, 582 (1994)  and references
therein.

\bibitem{nmr} G.~ Q.~ Zheng, T.~Odaguchi, Y.~ Kitaoka, K.~Asayama,
Y.~ Kodama, K.~Mizuhashi and S.~ Uchida S, {\sl Physica}
{\bf C 263}, 367 (1996).

\bibitem{kondo} J.~Kondo, {\sl Progr.~Theor.~Phys.\/}{\bf 32}, 37
(1964).

\bibitem{pwa} P.~W.~Anderson, {\sl J.~Phys.~C}~{\bf 3}, 2436 (1970).

\bibitem{nozieres}
P.~Nozi{\`e}res and A.~Blandin, {\sl J.~Physique}~{\bf 41}, 193 (1980).

\bibitem{wilson} K.~G.Wilson, {\sl Rev.~Mod.~Phys.\/} {\bf 47}, 773
(1975); H.~R.~Krishnamurthy, J.~Wilkins and K.~G.Wilson,
{\sl Phys.~Rev.~\/} {\bf B21}, 1044 (1980).

\bibitem{andrei} N.~Andrei, {\sl Phys.~Rev.~Lett.}~{\bf 45}, 379 (1980).

\bibitem{wiegmann} P.~B.~Wiegmann, {\sl JETP Lett.\/}{\bf 31}, 367
(1980).

\bibitem{read} N.~Read and D.~M.~Newns, {\sl J.~Phys.~C}~{\bf 16},
3273 (1983).

\bibitem{abgkv} A.~A.~Abrikosov and L.~P.~Gorkov, {\sl Sov.~Phys.~JETP}
{\bf 12}, 1243 (1961).

\bibitem{shiba} H.~Shiba, {\sl Progr.~Theor.~Phys.\/}{\bf 40}, 435
(1968).

\bibitem{fisher} Matthew P.~A.~ Fisher and Eduardo Fradkin,
{\sl Nucl. ~Phys.\/} {\bf B241 (FS13)}, 457 (1985).

\bibitem{dirac} Eduardo Fradkin,
{\sl Phys.~ Rev.\/} {\bf B 33}, 3257 (1986); {\sl Phys.~ Rev.~} {\bf B
33}, 3263 (1986).

\bibitem{ludwig} A.~ W.~ W.~Ludwig, M.~ P.~ A.~  Fisher, R.~  Shankar
and G.~ Grinstein,  {\sl Phys.~ Rev.\/} {\bf B 50}, 7526 (1994).

\bibitem{palee}P.~ A.~Lee,  {\sl Phys.~Rev.~Lett.}~{\bf 71}, 1887 (1993).

\bibitem{hirshfeld1} P.~J.~Hirschfeld and N.~Goldenfeld,
{\sl Phys.~ Rev.\/} {\bf B 48}, 4219 (1993).

\bibitem{hirshfeld2} K.~Ziegler, M.~Hettler and P.~J.~Hirschfeld,
{\sl Phys.~Rev.~Lett.}~{\bf 77}, 3013 (1996).

\bibitem{hirshfeld3}L .~S.~ Borkowski and P.~J.~Hirschfeld,
{\sl J.~ Low Temp.~ Phys.\/} {\bf 96},  185 (1994);
{\sl Phys.~ Rev.\/} {\bf B 46}, 9274 (1992).

\bibitem{kotliar} V.~Dobrosavljevic, T.~R.~Kirkpatrick  and G.~Kotliar,
{\sl Phys.~Rev.~Lett.~\/} {\bf 69}, 1113 (1992);
 E.~Miranda, V.~Dobrosavljevic and G.~Kotliar,
{\sl Phys.~Rev.~Lett.~\/} {\bf 78}, 290 (1997) and
{\sl Proceedings of the ITP-Santa Barbara conference on non-Fermi
 liquids}, cond-mat/9612160.

\bibitem{phillips}
I.~Martin, Yi Wan and P.~Phillips,
{\sl Phys.~Rev.~Lett.~\/} {\bf  78}, 114 (1997).

\bibitem{withoff} D.~Withoff and E.~Fradkin,
{\sl Phys.~Rev.~Lett.}~{\bf 64},
1835 (1990); D.~J.~Withoff, {\sl Path Integral Methods for the
large-N Kondo
Model}, University of Illinois Thesis, 1988.

\bibitem{jay} K.~Chen and C.~Jayaprakash, {\sl Phys.~ Rev.\/} {\bf B 52},
 14436 (1995); {\sl J.~Phys.~, Condensed Matter}{\bf 37},
L491-8(1995).

\bibitem{ingersent} K.~Ingersent, {\sl Phys.~Rev.~\/} {\bf B 54}, 11936
(1996); see also  K.~Ingersent in {\sl Proceedings of Physical
Phenomena at High Magnetic Fields-II}, Z.~Fisk, L.~ Gorkov, D.~Meltzer and
J.~R.~Schrieffer editors, p.~179 (1996).

\bibitem{paperII}
C.~Cassanello and E.~Fradkin, {\sl Phys.~Rev.~\/} {\bf B53}, 15079 (1996).

\bibitem{paper4} C.~Cassanello and E.~Fradkin, in preperation (1997).

\bibitem{scalapino} See for example the review by D.~J.~Scalapino,
{\sl Phys.~Rep.}~{\bf 250}, 329 (1995).

\bibitem{schrieffer} J.~R.~Schrieffer, {\sl Theory of Superconductivity},
Addison-Wesley, Redwood City (1983).


\bibitem{wollman} D~ A.~Wollman, D.~J.~  Van Harlingen, J.~ Giapintzakis
and D.~M.~ Ginsberg ,{\sl Phys.~Rev.~Lett.}~{\bf 74}, 797 (1995);
D.~J.~ Van Harlingen, D~ A.~Wollman, D.~M.~ Ginsberg and  A.~J.~Leggett,
{\sl Physica}{\bf C}, 122 (1994).

\bibitem{shen} Z.~X.~Shen {\it et.al.,\/} {\sl Phys.~Rev.~Lett.}~{\bf 70},
1553 (1993);
Z.~X.~Shen and D.~S.~Dessau, {\sl Phys.~Rep.}~{\bf 253}, 1 (1995);
Z.~X.~Shen {\it et.al.,\/} {\sl Science}~{\bf 267}, 343 (1995).

\bibitem{campuzano} H.~Ding {\it et.al.,\/} {\sl Phys.~Rev.~Lett.\/}
{\bf 74}, 2784 (1995);
M.~Norman {\it et.al.,\/} {\sl Phys.~Rev.}~{\bf B 52}, 615 (1995);
H.~Ding {\it et.al.,\/} cond-mat 9603044.

\bibitem{gross} D.~Gross and A.~Neveu, {\sl Phys.~Rev.~\/} {\bf D10},
3235 (1974).

\bibitem{book} E.~Fradkin, {\sl Field Theories of Condensed Matter
Systems}, Addison-Wesley, Redwood City (1991).


\bibitem{doug} N.~E.~Bickers, D.~J.~Scalapino and S.~White,
{\sl Phys.~Rev.~Lett.~\/} {\bf 62}, 961 (1989).



\bibitem{doniach} S.~Doniach and E.~H.~Sondheimer,
{\sl Green's Functions for Solid State Physicists},
Adison-Wesley, Redwood City, (1982).

\bibitem{newns} D.~M.~Newns and A.~C.~Hewson, {\sl J.~Phys.}~{\bf F 10},
2429 (1980).


\end{references}
\end{document}